\documentclass[a4paper,12pt]{article}
\pdfoutput=1

\usepackage{jheppub}
\usepackage[utf8]{inputenc}
\usepackage{amsmath,amssymb,amsthm}
\usepackage{physics,mathtools}
\usepackage{tikz,float}
\usepackage{bbm}
\usepackage{subcaption}
\usepackage{booktabs}
\usepackage{pifont}

\makeatletter
\def\@fpheader{\vspace{0.1mm}}
\makeatother
\newcommand{\bw}{\begin{widetext}}
\newcommand{\ew}{\end{widetext}}
\newcommand{\bea}{\begin{eqnarray}}
\newcommand{\eea}{\end{eqnarray}}
\newcommand{\be}{\begin{equation}}
\newcommand{\ee}{\end{equation}}
\newcommand{\bca}{\begin{cases}}
\newcommand{\eca}{\end{cases}}

\newcommand{\ben}{\begin{enumerate}}
\newcommand{\een}{\end{enumerate}}

\usetikzlibrary{patterns}
\makeatletter
\let\over\@@over
\makeatother

\title{{{Spin-resolved double-trace thermal coefficients in holography}}}
\author{Ilija Buri\'c,}
\emailAdd{burici@tcd.ie}
\author{Ivan Gusev,}
\emailAdd{gusevi@tcd.ie}
\author{Andrei Parnachev}
\emailAdd{parnachev@maths.tcd.ie}

\affiliation{School of Mathematics and Hamilton Mathematics Institute, Trinity College, Dublin 2, Ireland}

\abstract{
It was previously shown that the stress-tensor sector of the OPE, together with the KMS condition, fixes holographic thermal two-point functions at vanishing spatial separation. We extend this construction to nonzero spatial separation, where the KMS condition leaves a residual ambiguity that depends only on the spatial separation. We show that this ambiguity is fixed by the zero-frequency bulk wave equation, which can be solved analytically in terms of Heun functions. This gives an efficient method for computing thermal coefficients of double-trace operators resolved by spin. We also study the Lorentzian analytic structure of the resulting correlator and show that complex bulk-cone singularities which appear at spacelike separation in the stress-tensor sector do not persist in the full two-point function; they are resolved by the double-trace contribution.
}

\begin{document}

\maketitle

\section{Introduction}
\label{S:Introduction}

The thermal bootstrap is a variant of the conformal bootstrap\footnote{
Early works include \cite{Ferrara:1973yt,Polyakov:1974gs,Mack:1976pa}, while some recent applications include numerical studies of vector models \cite{Rattazzi:2008pe,El-Showk:2012cjh,El-Showk:2014dwa,Kos:2016ysd}
and analytical results on large-spin anomalous dimensions \cite{Fitzpatrick:2012yx,Komargodski:2012ek}.
See \cite{Simmons-Duffin:2016gjk,Poland:2018epd,Bissi:2022mrs,Rychkov:2023wsd}
for recent reviews.}
that aims to constrain, and in favorable cases determine, finite-temperature observables in conformal field theories (CFTs) using basic consistency conditions. The most studied observable within this approach has been the thermal two-point function, which is constrained by the combination of a thermal conformal-block decomposition and the Kubo-Martin-Schwinger (KMS) condition \cite{El-Showk:2011yvt,Iliesiu:2018fao,Marchetto:2023xap}.
\smallskip

In this paper, we focus on the theory dual to a scalar field $\Phi$ minimally coupled to Einstein gravity. At leading order in the large central charge, the thermal two-point function of the field $\phi$ dual to $\Phi$ receives contributions from two families of operators, the composites of the stress tensor and the double-trace operators constructed from $\phi$. Thermal coefficients of multi-stress-tensors can be obtained by a perturbative solution of the bulk wave equation, \cite{Fitzpatrick:2019zqz}. They have been used to derive the {\it bouncing singularities} \
\cite{Fidkowski:2003nf,Festuccia:2005pi} of the retarded two-point function in the complex time domain, \cite{Ceplak:2024bja}. \footnote{For more recent work on bouncing singularities, see \cite{Parisini:2023nbd,Afkhami-Jeddi:2025wra,Ceplak:2025dds,Dodelson:2025jff,Jia:2025jbi,AliAhmad:2026wem,Jia:2026pmv,Araya:2026shz,Giombi:2026kdz,Arnaudo:2026der,Grozdanov:2026cut,Jia:2026ryl,Arnaudo:2026tcy,Grozdanov:2026ktq,Dodelson:2026gak,Buric:2026qsp}.}
By contrast, the double-trace thermal coefficients are much more difficult to obtain from the bulk and numerical approaches based on the wave equation can only give the lowest-lying such coefficient, \cite{Parisini:2023nbd}. In \cite{Buric:2025anb,Buric:2025fye}, we developed an alternative method, based on the thermal bootstrap, for computing double-trace coefficients. This was applied to robustly obtain a number of low-lying coefficients. (See \cite{Niarchos:2025cdg,Barrat:2025twb,Arnaudo:2026der,Niarchos:2026wsw} for more work on double-trace thermal coefficients
and \cite{Iliesiu:2018zlz,Barrat:2025wbi,Barrat:2025nvu} for applications of the thermal bootstrap to other models.)
\smallskip

The works \cite{Buric:2025anb,Buric:2025fye} were concerned with the thermal two-point function with vanishing spatial separation between the two insertion points. Consequently, the obtained CFT data were weighted sums of thermal coefficients of families of operators that have different spins and the same scaling dimension.
In this paper, we will go a step further and determine thermal coefficients of individual double-trace operators.
\smallskip

Our analysis is based on two main ingredients. Starting from the stress-tensor sector of the two-point function, $g_\beta^T(\tau,x)$, which is obtained by the bulk analysis of \cite{Fitzpatrick:2019zqz}, we will perform a sum over images. This sum is divergent and we shall describe how to regularize it using analytic continuation. Once this is done, the resulting function, which is KMS-invariant by construction, is still not the holographic thermal two-point function and differs from the latter in its zero Matsubara mode. Our second step is to obtain the zero mode by solving a two-variable partial differential equation in the bulk. This is substantially easier than the three-variable wave equation that was analyzed in previous works, \cite{Parisini:2023nbd,Buric:2025fye}. Using the relation of the two-variable PDE to the Heun connection problem (see \cite{Aminov:2020yma,Dodelson:2022yvn,Aminov:2023jve,Arnaudo:2024rhv,Arnaudo:2024sen,Bajc:2025jjv} and references therein for similar observations) we will derive an exact solution in momentum space. This in turn gives the position-space solution to any desired precision. Determination of the zero mode fully fixes the thermal two-point function and consequently the double-trace coefficients.
\smallskip

While the procedure we described holds for every value of the scaling dimension $\Delta_\phi$, we carry out explicit numerical calculations for $\Delta_\phi = 3/2$. Some of the low-lying double-trace coefficients are given in equation \eqref{main-numbers}. Our findings agree with and considerably extend recent results of \cite{Niarchos:2026wsw}.
\smallskip

A further motivation for reconstructing the full two-point function is its Lorentzian
analytic structure. As shown in \cite{Araya:2026shz}, after analytic continuation to
Lorentzian time the stress-tensor sector by itself can develop singularities at real
values of $t$ with $|t|<x$, i.e. at spacelike separation. Such singularities are not
expected to be a property of the complete local thermal correlator, and it is natural to
ask how they are modified once the double-trace sector is included. In
Section \ref{S:Analytic structure in real time}, we show that these would-be bulk-cone
singularities do not persist in the full two-point function; they are canceled by the
double-trace sector.
\smallskip

The paper is organized as follows. In Section \ref{S:Two-point function at finite x separation}, we explain how the thermal two-point function, up to an undetermined zero Matsubara mode, is obtained from its stress-tensor sector by making use of the sum over images. The zero mode is computed in Section \ref{S:Zero mode from the bulk PDE} by solving a two-variable PDE in the bulk. In this section, we also present our main new results for double-trace thermal coefficients. Section \ref{S:Analytic structure in real time} is dedicated to singularities of the two-point function. We discuss our results in Section \ref{S:Discussion}.

\section{Two-point function at finite \texorpdfstring{$x$}{x} separation}
\label{S:Two-point function at finite x separation}

In this section, we describe how the thermal two-point function in holography is obtained from its stress-tensor sector. Our method consists of two steps: the (regularized) sum over images and a computation of the zero Matsubara mode. The second step and numerical results for double-trace coefficients are given in the next section.

\subsection{Conformal block decomposition}

We consider the Euclidean thermal two-point function at finite temporal and spatial separations. Rotational invariance implies that it depends only on $\tau$ and $x \equiv |\vec{x}|$. Its conformal block decomposition is
\begin{equation}\label{conformal-block-decomposition}
    g_\beta(\tau,x) \equiv \langle \phi(\tau,x) \phi(0,0) \rangle = \sum_{\Delta, J} 
a_{\Delta, J} \, C_J^{\left(1\right)} \!\left( \frac{\tau}{\sqrt{\tau^{2} + x^2}} \right)
\left( \tau^2 + x^2 \right)^{\frac{\Delta - 2 \Delta_{\phi}}{2}}\ .
\end{equation}
Here and throughout the section, we set $\beta=1$. Occasionally, in the analysis that follows, we will use variables other than $\tau$ and $x$. We shall denote
\begin{align}\label{change-of-vars-1}
    & z = \tau + ix = re^{i\theta}\,, \quad \bar{z} = \tau - ix = re^{-i\theta}\,, \\
    & r^2 = \tau^2 + x^2\,, \hskip1.3cm \eta = \frac{\tau}{r} = \cos\theta\ .\label{change-of-vars-2}
\end{align}
In holographic theories at leading order in $1/N$, the $\phi\times\phi$ OPE contains two families of operators - the composites of the stress tensor, called the multi-stress-tensor operators, and double-trace operators made out of $\phi$. Correspondingly, the two-point function \eqref{conformal-block-decomposition} decomposes into stress-tensor and double-trace sectors,
\begin{equation}
    g_\beta(\tau,x) = g_\beta^T(\tau,x) + g_\beta^{\rm DT}(\tau,x)\ .
\end{equation}
Double-trace operators of spin $J$ and containing ‘‘$m$ boxes" will be denoted by
\begin{equation}
    [\phi\phi]_{m,J}\,, \qquad \Delta_{m,J}=2\Delta_\phi+2m+J\,,
\end{equation}
and the corresponding thermal coefficients by $b_{m,J}$. With this notation, the double-trace sector takes the form
\begin{equation}\label{double-trace-CB-decomposition}
    g_\beta^{\rm DT}(\tau,x)
    = \sum_{m=0}^\infty \sum_{J=0,2,\ldots}^\infty b_{m,J}\, r^{2m+J} C_J^{(1)}(\eta) 
    = \sum_{q=0}^\infty r^{2q} \sum_{J=0,2,\ldots,2q} b_{q-J/2,J}\, C_J^{(1)}(\eta)\ .
\end{equation}
In the following sections, we will find the monomial form of the double-trace sector expansion particularly useful
\begin{equation}
g_\beta^{\rm DT}(\tau,x) = \sum_{k=0}^{\infty} \sum_{i=0}^{\infty} a_{k,i}\, x^{2k}\tau^{2i}\ . \label{app:dt-monomial-expansion}
\end{equation}
The change of basis from the monomial coefficients $a_{k,i}$ to the double-trace thermal coefficients $b_{m,J}$ is
\begin{equation}\label{double-trace-from-expansion-Section2}
    b_{m,J} = \sum_{k=0}^{m+J/2} a_{k,\,m+J/2-k}\, \mathcal M^{(m,J)}_k\,,
\end{equation}
where the matrix elements (see Appendix \ref{app:double-trace-change-of-basis}) are
\begin{equation}\label{double-trace-from-expansion-matrix}
\mathcal M^{(m,J)}_k = \frac{2}{\pi}\sum_{r=0}^{J/2}(-1)^r 2^{J-2r} \binom{J-r}{r}B\!\left(m+J-k-r+\frac12,\,k+\frac32\right)\ .
\end{equation}
Similarly, the multi-stress-tensors carry a pair of labels $(n,J)$, where $J$ is the spin and the scaling dimension of the operator equals $\Delta=4n$. We denote the corresponding thermal coefficients by $\lambda_{n,J}$, so that the conformal block decomposition of the stress-tensor sector reads
\begin{equation}\label{gT-conformal-blocks}
    g_\beta^T(\tau,x)=\frac{1}{r^{2\Delta_\phi}}\sum_{n=0}^\infty\sum_{J=0,2,\ldots,2n}\!\!\!\lambda_{n,J}\,r^{4n}C_{J}^{(1)}\left(\eta\right)\ .
\end{equation}
For the theory dual to a scalar field minimally coupled to Einstein gravity, the stress-tensor sector is obtained by solving the Klein-Gordon equation on the planar AdS black hole in a near-boundary expansion, \cite{Fitzpatrick:2019zqz}. This gives another expansion of the stress-tensor sector,
\begin{equation}\label{e.fgege}
    g_\beta^T(\tau,x) = \frac{1}{r^{2\Delta_\phi}} \sum_{i=0}^\infty \sum_{j=0}^i a^i_{j,2i-j} r^{4i-2j} x^{2j}\ .
\end{equation}
Our notation follows \cite{Buric:2025fye}, where details about how coefficients $a^i_{j,2i-j}$ are computed can be found. The change of basis which expresses coefficients $\lambda_{n,J}$ in terms of the $a^i_{j,2i-j}$ is derived in Appendix \ref{app:stress-tensor-change-of-basis}:
\begin{equation}
    \lambda_{n,J} = \sum_{j=0}^{n} \mathcal{M}_{jJ}\,a^n_{j,2n-j}\,,
    \qquad
    J=0,2,\ldots,2n\,,
\end{equation}
where
\begin{equation}
    \mathcal{M}_{jJ} \equiv \frac{(-1)^{J/2}4^{-j}\Gamma(2j+2)}{\Gamma(j+1-J/2)\Gamma(j+2+J/2)}\ .
\end{equation}

\subsection{Two-point function reconstruction}

In this subsection, we describe how to reconstruct the double-trace sector of the two-point function from the stress-tensor sector. In coordinates, $(\tau,x)$, the latter is given by
\begin{equation}\label{stress-tensor_g}
    g_\beta^T(\tau, x) = \frac{1}{(\tau^2 + x^2)^{\Delta_\phi}}\sum_{i=0}^\infty \sum_{j=0}^i a_{j,\,2i-j}^{i}\, x^{2j}(\tau^2 + x^2)^{2i-j}\ .
\end{equation}
To construct the full two-point function, we will make use of the method of images and try to compute
\begin{equation}\label{method-of-images}
    g^{\text{images}}_\beta(\tau,x) = \sum_{m=-\infty}^\infty g_\beta^T(\tau+m,x)\ .
\end{equation}
Upon expanding $g_\beta^T(\tau+m,x)$ according to \eqref{e.fgege} and exchanging the order of summations, the resulting image sums diverge and need to be regularized. We will do this by analytic continuation which makes use of Hurwitz $\zeta$--functions. After a further Pad\'e-Borel resummation, the resulting function is well-defined and KMS invariant.\footnote{For half-integer $\Delta_\phi$, additional care is needed, as we explain in detail below.} However, it is not yet the two-point function we are after. Indeed, we note that any function of the form
\begin{equation}\label{KMS-invariant-shift}
     g^{\text{images}}_\beta(\tau,x) + f(x)\,,
\end{equation}
is KMS invariant, decomposes into stress-tensor and double-trace sectors and has the stress-tensor sector given by \eqref{stress-tensor_g}. The actual two-point function takes the form
\begin{equation}
    g_\beta(\tau,x) = g^{\text{images}}_\beta(\tau,x) + c(x)\,,
\end{equation}
for an appropriate function $c(x)$. One way of fixing $c(x)$ is to supply the zero mode of the Fourier decomposition of the two-point function. Let us write the Fourier decomposition as
\begin{equation}
    g_\beta(\tau,x) = \sum_{n=0}^\infty g^{(n)}(x) \cos(2\pi n \tau) \equiv g^{(0)}(x) + g^{(n\neq0)}(\tau,x) \ .
\end{equation}
The second term $g^{(n\neq0)}(\tau,x)$ is invariant under shifts \eqref{KMS-invariant-shift} and thus fully determined by the function $g^{\rm images}_\beta(\tau,x)$ that we have obtained above. Schematically,
\begin{equation}\label{projection}
    g^{(n\neq0)}(\tau,x) = \Pi \cdot g^{\text{images}}_\beta(\tau,x)\,,
\end{equation}
where $\Pi$ is the projection to non-zero Fourier modes. On the other hand, the zero mode $g^{(0)}(x)$ will be obtained by solving a two-variable PDE in the bulk. The point here is that the original three-variable bulk equation has a well-defined reduction to the zero mode subspace. Having obtained both $g^{(0)}(x)$ and $g^{(n\neq0)}(\tau,x)$, we will read off the OPE coefficients.
\smallskip

Finally, let us mention an additional subtlety that we skipped over in order to have a clear presentation. We solve the bulk PDE for particular values of $\Delta_\phi=3/2,5/2$, for which the boundary conditions are most easily imposed. For concreteness, we shall focus on $\Delta_\phi=3/2$. This is precisely one of the non-generic values of $\Delta_\phi$ for which the construction of $g^{\text{images}}_\beta(\tau,x)$ by analytic continuation suffers from singularities. However, we shall see that the singularities are only present in the zero mode and thus disappear in the projection \eqref{projection}. We shall therefore first derive a formula for thermal coefficients at generic $\Delta_\phi$ and take the limit $\Delta_\phi\to3/2$ at the end. This limit is well-defined and gives finite thermal coefficients. We proceed to carry out each of the above steps.

\subsection{Analytic continuation of the image sum}

To understand how to perform the sum over images, it is useful to first consider the simpler case of the generalized free field
\begin{equation}\label{GFF2PF}
g^{\rm GFF}_\Delta(\tau,x) = \sum_{m\in\mathbb{Z}} \frac{1}{\big[(\tau+m)^2+x^2\big]^\Delta}\ .
\end{equation}
To analytically continue from $\Delta>1/2$, for which the sum converges, we assume $x < |\tau + m|$, expand the denominator for each fixed $m$ using the binomial theorem and exchange the order of summations to get
\begin{align}\label{GFF-Hurwitz-zeta-summation}
g^{\rm GFF}_\Delta(\tau,x) 
&= \sum_{m\in\mathbb{Z}}\sum^\infty_{q=0}\binom{-\Delta}{q} x^{2q} |\tau+m|^{-2\Delta-2q}\\
&= \sum^\infty_{q=0}\binom{-\Delta}{q} x^{2q} \left( \zeta_H(2q+2\Delta,\tau) + \zeta_H(2q+2\Delta,1-\tau)\right)\ . \nonumber
\end{align}
Note that this expression contains singular terms when $\Delta$ is a half-integer with $\Delta\leq1/2$. We will discuss this further below. For now, we keep $\Delta$ generic and apply \eqref{GFF-Hurwitz-zeta-summation} term by term to the stress-tensor sector \eqref{stress-tensor_g}. This gives the following representation of the image sum \eqref{method-of-images}
\begin{align}\label{2PF_nz}
    g^{\rm images}_\beta(\tau,x) = & \sum_{i=0}^\infty \sum_{j=0}^i \sum^\infty_{q=0}a_{j,\,2i-j}^i\, \binom{2i-j-\Delta_\phi}{q} x^{2j+2q}\\
    \times & \left( \zeta_H(2j + 2q -4i +2\Delta_\phi,\tau) + \zeta_H(2j + 2q -4i +2\Delta_\phi,1-\tau ) \right)\ . \nonumber
\end{align}
To simplify expressions, we will denote the symmetric combination of Hurwitz zeta functions by
\begin{equation}
Z(s,\tau)=\zeta_H(s,\tau)+\zeta_H(s,1-\tau)\ .
\end{equation}
Furthermore, introducing $k = j + q$, we can rewrite \eqref{2PF_nz} as
\begin{equation}\label{zeta-regularised-image-sum}
    g^{\rm images}_\beta(\tau,x) = \sum_{i=0}^\infty \sum_{j=0}^i \sum^\infty_{k=j} a_{j,\,2i-j}^{i}\, \binom{2i-j-\Delta_\phi}{k-j} x^{2k} Z(2k-4i+2\Delta_\phi,\tau)\ .
\end{equation}
\paragraph{Expansion coefficients.} In the rest of our analysis, we shall work directly with expansion coefficients of $g^{\rm images}_\beta(\tau,x)$, rather than the function itself (these are related to thermal coefficients of double-trace operators by \eqref{double-trace-from-expansion-Section2}). The representation \eqref{zeta-regularised-image-sum} is convenient for this purpose. Let us fix a $k\geq0$. The coefficient of $x^{2k}$ in the expansion of $g^{\rm images}_\beta$ reads
\begin{equation}\label{x2k-1st-expression}
    \left. g_\beta^{\rm images} \right|_{x^{2k}}  = \sum_{j=0}^k \sum_{i=j}^\infty a_{j,\,2i-j}^{i}\, \binom{2i-j-\Delta_\phi}{k-j}Z(2k-4i+2\Delta_\phi,\tau)\ .
\end{equation}
Next, we expand the last expression in $\tau$. Using the binomial expansion
\begin{equation}
(n\pm \tau)^{-s} = n^{-s}\left(1\pm\frac{\tau}{n}\right)^{-s} = \sum_{k=0}^\infty \binom{-s}{k}(\pm 1)^k \frac{\tau^k}{n^{s+k}}\,,  \qquad |\tau|<1\,, \quad n\neq0\,,
\end{equation}
and summing over \(n\geq 1\) we obtain
\begin{equation}
\sum_{n=1}^\infty (n\pm \tau)^{-s} = \sum_{k=0}^\infty \binom{-s}{k}(\pm 1)^k \tau^k \,\zeta(s+k)\ .
\end{equation}
Including the \(n=0\) term \(\tau^{-s}\) from \(\zeta_H(s,\tau)\) gives
\begin{equation}
Z(s,\tau)=\tau^{-s}+\sum_{k=0}^\infty \binom{-s}{k}\big(1+(-1)^k\big)\tau^k \zeta(s+k) = \tau^{-s}+2\sum_{l=0}^\infty \binom{-s}{2l}\,\zeta(s+2l)\,\tau^{2l}\ .
\end{equation}
One should be careful with the case where $s+2l=1$: the pole of the $\zeta$ function is canceled by the zero of the binomial coefficient. Substituting back into \eqref{x2k-1st-expression}, the coefficient of $x^{2k}$ takes the form
\begin{align}
    & \left. g_\beta^{\rm images} \right|_{x^{2k}} =  \sum_{j=0}^k \sum_{i=j}^\infty a_{j,\,2i-j}^{i}\, \binom{2i-j-\Delta_\phi}{k-j}\tau^{4i-2k-2\Delta_\phi} \\ \nonumber
    &+ 2\sum_{j=0}^k \sum_{i=j}^\infty\sum_{l=0}^\infty a_{j,\,2i-j}^{i}\, \binom{2i-j-\Delta_\phi}{k-j}\binom{4i-2k-2\Delta_\phi}{2l}\zeta(-4i+2k+2\Delta_\phi+2l)\tau^{2l}\ .
\end{align}
As one can see from this formula, the expression is naturally decomposed into the stress-tensor sector (the first line) and the double-trace sector (the second line). We proceed to analyze the two sectors in turn.

\paragraph{The stress-tensor sector.} Expansion coefficients are given by finite sums
\begin{equation}\label{stress-tensor coefs}
    \left. g_\beta^{\rm images} \right|_{x^{2k}\tau^{4i-2k-2\Delta_\phi}} = \sum_{j=0}^{\text{min}(k,i)} a_{j,\,2i-j}^{i}\, \binom{2i-j-\Delta_\phi}{k-j}\ .
\end{equation}
These are nothing else but the expansion coefficients of the function $g_\beta^T(\tau,x)$.

\paragraph{The double-trace sector.} Expansion coefficients are given by
\begin{align}\label{double-trace coefs}
    &\left. g_\beta^{\rm images} \right|_{x^{2k}\tau^{2l}} = \\
    & 2\sum_{j=0}^k \sum_{i=j}^\infty a_{j,\,2i-j}^{i}\, \binom{2i-j-\Delta_\phi}{k-j}\binom{4i-2k-2\Delta_\phi}{2l}\zeta(-4i+2k+2\Delta_\phi+2l)\ .\nonumber
\end{align}
Unlike in the stress-tensor sector, the sum is infinite. Moreover, the series is asymptotic since
\begin{equation}
    \zeta(-4i+2k+2\Delta_\phi+2l) \sim (4i)!  
\end{equation}
have factorial, Gevrey-4 growth. The series can be regularized using Pad\'e-Borel resummation, similar to \cite{Buric:2025fye}. We denote the coefficients by
\begin{equation}\label{coefficients-aki-i>0}
    a_{k,l} = \int_0^\infty dt\, e^{-t}\, \text{Pad\'e}
    \left[\mathcal{B}_4 \left( \left. g_\beta^{\rm images} \right|_{x^{2k}\tau^{2l}} \right)\right]\,, \qquad l>0\,,
\end{equation}
where the Borel transform is defined as
\begin{align}\label{coefficients-aki-Borel}
    &\mathcal{B}_4\left( \left. g_\beta^{\rm images} \right|_{x^{2k}\tau^{2l}} \right) = \\
    & 2\sum_{j=0}^k \sum_{i=j}^\infty \frac{a_{j,\,2i-j}^i}{(4i)!}\, \binom{2i-j-\Delta_\phi}{k-j}\binom{4i-2k-2\Delta_\phi}{2l}\zeta(-4i+2k+2\Delta_\phi+2l)\,t^{4i}\ . \nonumber
\end{align}
In practice, the infinite sum over $i$ in the second line is truncated at order $N$, which is determined by the number of available coefficients $a_{j,\,2i-j}^i$. Furthermore, the resulting polynomial of degree $N$ in $t^4$ is replaced by its diagonal Pad\'e approximant before evaluating the integral \eqref{coefficients-aki-i>0}. For half-integer $\Delta_\phi$, zeta functions in \eqref{double-trace coefs} may exhibit singularities. However, for all $l>0$, the singular term comes multiplied by a vanishing binomial coefficient and their product is well-defined and finite.

\subsection{Projecting out the zero mode}

In the previous subsection, we considered the expansion of the $\zeta$-regularized image sum $g_\beta^{\rm images}(\tau,x)$ in powers of $\tau^2$ and $x^2$. Grouping the terms according to the power of $\tau^2$, we may write
\begin{equation}\label{g-images-tau-powers}
    g_\beta^{\rm images}(\tau,x) = g_{\rm images}^{[0]}(x) + g^{[1]}(x) \tau^2 + g^{[2]}(x) \tau^4 + \dots + g^T_\beta(\tau,x)\ .
\end{equation}
For the case of interest $\Delta_\phi = 3/2$, all terms except for $g_{\rm images}^{[0]}(x)$ are well defined. As we have argued at the beginning of the section, the true holographic two-point function is given by
\begin{equation}\label{g-true-tau-powers}
    g_\beta(\tau,x) = g^{[0]}(x) + g^{[1]}(x) \tau^2 + g^{[2]}(x) \tau^4 + \dots + g^T_\beta(\tau,x)\,,
\end{equation}
for some function $g^{[0]}(x)$. Notice that functions multiplying powers $\tau^{2n}$ with $n>0$ are the same in \eqref{g-images-tau-powers} and \eqref{g-true-tau-powers}. We are after determining $g^{[0]}(x)$. To this end, it is useful to consider one more decomposition of $g_\beta(\tau,x)$ - that in Fourier modes. Indeed, since $g_\beta(\tau,x)$ is periodic in $\tau$, it may be expanded in Fourier series
\begin{equation}
    g_\beta(\tau,x) = g^{(0)}(x) + g^{(1)}(x) \cos(2\pi\tau) + g^{(2)}(x) \cos(4\pi\tau) + \dots \ .
\end{equation}
Likewise, we can decompose $g_\beta^{\rm images}(\tau,x)$ in Fourier series. To achieve this, we pass to yet another representation of the GFF two-point function \eqref{GFF2PF},
\begin{equation}\label{Bessel GFF}
g^{\rm GFF}_{\Delta}(\tau,x) = \frac{\sqrt{\pi}\,\Gamma\!\left(\Delta-\frac12\right)}{\Gamma(\Delta)}x^{1-2\Delta}
+\frac{4\sqrt{\pi}}{\Gamma(\Delta)}\sum_{n=1}^\infty\left(\frac{x}{\pi n}\right)^{\frac12-\Delta}
K_{\frac12-\Delta}(2\pi n x)\, \cos(2\pi n \tau)\,,
\end{equation}
where $K_\nu(z)$ is a modified Bessel function of the second kind. This representation is valid for $x>|\text{Im}\tau|$. Details of its derivation can be found in Appendix \ref{app:gff_bessel}. We perform the sum over images of the stress-tensor part of the two-point function \eqref{stress-tensor_g} term by term in the conformal block decomposition and use the above analytic continuation to write
\begin{align}\label{g-images-K}
& g^{\rm images}_\beta(\tau,x) = \sum_{i=0}^\infty\sum_{j=0}^i
a^i_{j,2i-j} \frac{\sqrt{\pi}\,\Gamma\!\left(\Delta_\phi-2i+j-\frac12\right)}{\Gamma(\Delta_\phi-2i+j)}x^{1-2\Delta_\phi+4i} + \\
&\sum_{i=0}^\infty \sum_{j=0}^i a_{j,\,2i-j}^i\, x^{2j} \frac{4\sqrt{\pi}}{\Gamma(\Delta_\phi-2i+j)}
\sum_{n=1}^\infty \left(\frac{x}{\pi n}\right)^{\frac12-\Delta_\phi+2i-j} K_{\frac12-\Delta_\phi+2i-j}(2\pi n x)\, \cos(2\pi n \tau)\ . \nonumber
\end{align}
In their overlapping domain of convergence, the representation \eqref{g-images-K} is equal to the $\zeta$-function representation \eqref{zeta-regularised-image-sum} of $g^{\rm images}_\beta$. Crucially, \eqref{Bessel GFF} and \eqref{g-images-K} are already written in the form of a Fourier series. The zero Fourier mode of $g^{\rm images}_\beta$ is given by the first line in \eqref{g-images-K},
\begin{equation}\label{g-images-zero-mode}
    g^{(0)}_{\rm images}(x) = \sum_{i=0}^\infty\sum_{j=0}^i
a^i_{j,2i-j} \frac{\sqrt{\pi}\,\Gamma\!\left(\Delta_\phi-2i+j-\frac12\right)}{\Gamma(\Delta_\phi-2i+j)}x^{1-2\Delta_\phi+4i}\ .
\end{equation}
We are now ready to extract the function $g^{[0]}(x)$. Let us denote
\begin{equation}\label{c(x)-definition}
    c(x) \equiv g_\beta(\tau,x) - g_\beta^{\rm images}(\tau,x) \ .
\end{equation}
Since $c(x)$ does not depend on $\tau$, we can express it in two different ways
\begin{equation}
    c(x) = g^{[0]}(x) - g_{\rm images}^{[0]}(x) = g^{(0)}(x) - g^{(0)}_{\rm images}(x)\ .
\end{equation}
Therefore, we find
\begin{equation}\label{g[0]-expression}
    g^{[0]}(x) = g^{(0)}(x) + g^{[0]}_{\rm images}(x) - g^{(0)}_{\rm images}(x)\,,
\end{equation}
where the Taylor expansion of $g^{[0]}_{\rm images}(x)$ is given in \eqref{double-trace coefs} and $g^{(0)}_{\rm images}(x)$ is given in \eqref{g-images-zero-mode}. The final ingredient that we need to get $g^{[0]}(x)$, and thus double-trace coefficients, is the function $g^{(0)}(x)$. This is obtained by solving a two-variable PDE in the bulk in the next section. Finally, note that in the case $\Delta_\phi=3/2$ both $g^{[0]}_{\rm images}(x)$ and $g^{(0)}_{\rm images}(x)$ in \eqref{g[0]-expression} become singular. However, these singularities cancel each other to give a well-defined $g^{[0]}(x)$. We proceed to explain this in more detail.

\paragraph{Limit to $\Delta_\phi = \frac32$ and expansion coefficients.} In order to explain how the singularities present for $\Delta_\phi = 3/2$ get canceled in \eqref{g[0]-expression}, we keep $\Delta_\phi$ slightly away from this value and set
\begin{equation}
    \Delta_\phi=\frac32+\epsilon\ .
\end{equation}
For $l=0$, equation \eqref{double-trace coefs} gives
\begin{equation}\label{x2k_tau0_regulated}
    \left. g_\beta^{\rm images} \right|_{x^{2k}\tau^{0}} = 2\sum_{j=0}^k \sum_{i=j}^\infty
    a_{j,\,2i-j}^i\, \binom{2i-j-\frac32-\epsilon}{k-j} \zeta(-4i+2k+3+2\epsilon)\ .
\end{equation}
The only possible divergences come from the pole of the Riemann zeta function at unit argument. Thus, the singular terms appear for $i=\frac{k+1}{2}$. We can write the coefficients as
\begin{align}\label{singularity-zeta}
\left. g_\beta^{\rm images} \right|_{x^{2k}\tau^{0}}
&=
\frac{\delta_{k\,{\rm odd}}}{\epsilon}
\sum_{j=0}^{\frac{k+1}{2}}
a_{j,\,k+1-j}^{\frac{k+1}{2}}\,
\binom{k-j-\frac12}{k-j}
\nonumber\\[1mm]
& + 2\sum_{j=0}^k \sum_{\substack{i=j\\ i\neq \frac{k+1}{2}}}^\infty a_{j,\,2i-j}^i\,
\binom{2i-j-\frac32}{k-j}\,
\zeta(-4i+2k+3)\\[1mm]
&+ \delta_{k\,{\rm odd}}
\sum_{j=0}^{\frac{k+1}{2}}
a_{j,\,k+1-j}^{\frac{k+1}{2}}\,
\binom{k-j-\frac12}{k-j}
\left[2\gamma_{\rm E}-\psi\!\left(k-j+\frac12\right)
+\psi\!\left(\frac12\right)
\right]
+O(\epsilon)\,, \nonumber
\end{align}
where
\begin{equation}
    \delta_{k\,{\rm odd}} = \frac{1-(-1)^k}{2}\ .
\end{equation}
Here, $\gamma_{\rm E}$ is the Euler--Mascheroni constant and $\psi(z)=\Gamma'(z)/\Gamma(z)$ is the digamma function.
\smallskip

Consider now \eqref{g-images-zero-mode} in the same limit. We denote the coefficient of $x^{2k}$ of this term by $\left. g^{(0)}_{\rm images} \right|_{x^{2k}}$ and compute
\begin{equation}\label{singularity-Bessel}
\begin{aligned}
\left. g^{(0)}_{\rm images} \right|_{x^{2k}} &=\delta_{k\,{\rm odd}} \sum_{j=0}^{\frac{k+1}{2}}
a_{j,\,k+1-j}^{\frac{k+1}{2}}\, \binom{k-j-\frac12}{k-j} \\
&\qquad \times
\left[\frac{1}{\epsilon} -2\log x +\psi(k+1-j) -\psi\!\left(j-k+\frac12\right) \right]
+O(\epsilon)\ .
\end{aligned}
\end{equation}
Therefore, we see that the $O(1/\epsilon)$ terms cancel between \eqref{singularity-zeta} and \eqref{singularity-Bessel} to give a well-defined quantity in \eqref{g[0]-expression}. By taking the $\epsilon\to0$ limit, we find
\begin{align}\label{extracting-expansion-coefficients}
\nonumber
    \left. g^{[0]}(x) \right|_{x^{2k}} & = \left. g^{(0)}(x) \right|_{x^{2k}} + 2\sum_{j=0}^k
\sum_{\substack{i=j\\ i\neq \frac{k+1}{2}}}^{\infty}
a_{j,\,2i-j}^{i}\,
\binom{2i-j-\frac32}{k-j}\,
\zeta(-4i+2k+3)\\
& +\, \delta_{k\,{\rm odd}}
\sum_{j=0}^{\frac{k+1}{2}}
a_{j,\,k+1-j}^{\frac{k+1}{2}}\,
\binom{k-j-\frac12}{k-j}
\left[2\log x - 2\log 2 + \gamma_{\rm E} - \psi(k+1-j)\right]\ . 
\end{align}
Let us comment on the appearance of logs in the second line. Logarithms will also appear in the decomposition of the zero mode function $g^{(0)}(x)$ and our notation in \eqref{extracting-expansion-coefficients} is meant to indicate that the coefficient of $x^{2k}$ is of the form $\gamma_k + \tilde\gamma_k \log(x)$, where $\gamma_k,\tilde\gamma_k$ are numbers. The log terms will cancel between the first and second line, giving a pure series expansion for $g^{[0]}(x)$. Taking this into account, we can write \eqref{extracting-expansion-coefficients} as 
\begin{align}\label{extracting-expansion-coefficients-simpler}
\nonumber
    a_{k,0} & \equiv \left. g^{[0]}(x) \right|_{x^{2k}} = \gamma_k + 2\sum_{j=0}^k
\sum_{\substack{i=j\\ i\neq \frac{k+1}{2}}}^{\infty}
a_{j,\,2i-j}^i\,
\binom{2i-j-\frac32}{k-j}\,
\zeta(-4i+2k+3)\\
& +\, \delta_{k\,{\rm odd}}
\sum_{j=0}^{\frac{k+1}{2}} a_{j,\,k+1-j}^{\frac{k+1}{2}}\,\binom{k-j-\frac12}{k-j} \left[- 2\log 2 + \gamma_{\rm E} - \psi(k+1-j)\right]\ . 
\end{align}
Finally, we note that, according to \eqref{g-images-zero-mode}, $g^{(0)}_{\rm images}$ contains a term $2/x^2$ arising from the $i=j=0$ contribution. When computing $g^{[0]}(x)$ using \eqref{g[0]-expression}, this term precisely cancels the corresponding contribution to $g^{(0)}(x)$ given in \eqref{eq:finite-exact-position-space-correlator}.

\section{Zero mode from the bulk PDE}
\label{S:Zero mode from the bulk PDE}

The purpose of this section is to determine the zero-frequency correlator $g^{(0)}(x)$ by solving a two-dimensional PDE. The original problem reduces by one dimension after Fourier transforming to frequency space and setting $\omega=0$. The resulting solution determines the coefficients in \eqref{extracting-expansion-coefficients-simpler}, which contribute to the double-trace coefficients after the change of basis described in Appendix \ref{app:double-trace-change-of-basis}.

\subsection{Bulk PDE and boundary conditions}

The zero-frequency two-point function is obtained by solving the Klein--Gordon equation
\begin{equation}
\left(\Box-\Delta_\phi(\Delta_\phi-4)\right)\Phi=0\,,
\end{equation}
on the Euclidean planar black hole background with the metric
\begin{equation}
ds^2=\frac{1}{z^2}
\left(\frac{dz^2}{f(z)}+f(z)d\tau^2+d\vec{x}^{2}\right)\,,
\qquad f(z)=1-z^4\ .
\end{equation}
Our conventions are such that the horizon sits at $z_H=1$, so that the inverse temperature is $\beta=\pi$. Furthermore, we shall write $x=|\vec{x}|$. After Fourier transforming along the Euclidean-time direction and setting $\omega=0$, rotational invariance in the spatial directions reduces the Klein--Gordon equation to
\begin{equation}\label{dif-op}
\left[
z^5\partial_z\left(\frac{f(z)}{z^3}\partial_z\right)
+z^2\left(\partial_x^2+\frac{2}{x}\partial_x\right)
-\Delta_\phi(\Delta_\phi-4)
\right]\Phi_0(z,x)=0\ .
\end{equation}
The solution is required to reproduce a point-like source at the asymptotic boundary $z=0$ and to be regular at the Euclidean horizon $z=1$. In addition, it must be regular at $x=0$ away from the boundary source and decay as $x\to\infty$. These boundary conditions and details of the following discussion are presented in Appendix \ref{app: zero frequency}, see in particular equations \eqref{bc-s-Phi-appendix} and \eqref{H-bdy-conditions}-\eqref{H-horizon-condition}. For $\Delta_\phi=3/2$, we split the bulk field into two pieces,
\begin{equation}
    \Phi_0=G_{\rm tAdS}^{(0)}+z^{3/2}H\ .
\end{equation}
The first piece, $G_{\rm tAdS}^{(0)}$, is known analytically. It is the zero-frequency part of the thermal AdS bulk-to-boundary propagator. Explicitly,
\begin{equation}
    G_{\rm tAdS}^{(0)}(z,x)=\frac{2}{\pi}\frac{z^{3/2}}{z^2+x^2}\ .
\end{equation}
This term is introduced because it already contains the singular behavior associated with the coincident boundary source. In particular, near the boundary,
\begin{equation}
    G_{\rm tAdS}^{(0)}(z,x)\sim z^{3/2}\frac{2}{\pi x^2}\ .
\end{equation}
The remaining function $H$ is smoother and contains the thermal correction due to the black-hole geometry. Substituting this decomposition into the Klein--Gordon equation gives an inhomogeneous PDE for $H$. The source on the right-hand side appears because $G_{\rm tAdS}^{(0)}$ solves the scalar equation in pure AdS, but not in the black-hole background. In other words, the PDE for $H$ measures the difference between the black-hole geometry and pure AdS. The coefficient multiplying $z^{3/2}$ is the zero-frequency boundary two-point function. Therefore,
\begin{equation}
g^{(0)}(x)=\frac{2}{\pi x^2}+H(0,x)\ .
\end{equation}

\subsection{Momentum space and reduction to a one-variable problem}

In this and the next subsection, we explain how the Klein--Gordon equation from above may be exactly solved in terms of Heun functions. This subsection is dedicated to reducing the equation to a single-variable problem, while the solution is derived in the next.

To begin with, we consider the rescaled fields
\begin{equation}
\Psi(z,x) \equiv z^{-3/2}\Phi_0(z,x)\,, \qquad \widetilde{\Psi}_{\mathrm{tAdS}} \equiv z^{-3/2}G_{\rm tAdS}^{(0)}(z,x)\ .
\end{equation}
The equation for $\Psi$ reads
\begin{equation}\label{eq:Psi-zero-mode}
\left[(1-z^4)\partial_z^2 - 4z^3\partial_z - \frac94 z^2 +\partial_x^2+\frac{2}{x}\partial_x\right]\Psi(z,x)=0\ .
\end{equation}
We now Fourier transform along the three boundary spatial directions, using the convention
\begin{equation}
\Psi(z,\vec{x}) = \int\frac{d^3k}{(2\pi)^3} e^{i\vec{k}\cdot\vec{x}}\psi_k(z)\,, \qquad k\equiv |\vec{k}|\ .
\end{equation}
The Fourier modes $\psi_k(z)$ encode the momentum-space zero-frequency correlator as
\begin{equation}\label{gtilde-psi}
\widetilde{g}^{(0)}(k)=\psi_k(0)\ .
\end{equation}
Using the fact that $\Psi$ is rotationally invariant, the equation \eqref{eq:Psi-zero-mode} reduces to the ordinary differential equation for the Fourier modes
\begin{equation}\label{eq:radial-zero-frequency}
\left((1-z^4)\partial_z^2 - 4z^3\partial_z - \frac94 z^2 - k^2 \right) \psi_k(z) = 0\ .
\end{equation}
Let us discuss the boundary conditions satisfied by $\psi_k(z)$. In what follows, we shall denote differentiation with respect to $z$ by a prime. Regularity at the Euclidean horizon $z=1$ implies $\psi_k^{(n)}$ are finite, so
\begin{equation}\label{eq:horizon-regularity}
\psi_k'(1) = -\frac{k^2+\frac94}{4}\psi_k(1)\ .
\end{equation}
The spatial Fourier transform of the thermal-AdS contribution $\Psi(z,x)$ reads
\begin{equation}
\widetilde{\Psi}_{\mathrm{tAdS}}(z,k) = \frac{2}{\pi}\int d^3x\, \frac{e^{-i\vec{k}\cdot\vec{x}}}{x^2+z^2} = \frac{4\pi}{k}e^{-kz}\ .
\end{equation}
Therefore, the required condition at the conformal boundary $\{z=0\}$ is
\begin{equation}\label{eq:boundary-derivative-condition}
\psi_k'(0) = -4\pi + h_k'(0) = -4\pi\ .
\end{equation}
In the last line, $h_k(z)$ denotes the Fourier transform of $H(z,x)$ and we have used the boundary condition for $H$ given in \eqref{H-bdy-conditions}.

\subsection{Solution in terms of Heun functions}

We now solve the differential equation \eqref{eq:radial-zero-frequency} subject to boundary conditions \eqref{eq:horizon-regularity} and \eqref{eq:boundary-derivative-condition}. To this end, we set $t=z^2$. The equation \eqref{eq:radial-zero-frequency} becomes
\begin{equation}\label{eq:heun-zero-mode}
\frac{d^2\psi_k}{dt^2} +
\left(\frac{1}{2t}+\frac{1}{t-1}+\frac{1}{t+1}\right)\frac{d\psi_k}{dt}
+\frac{\frac{9}{16}t+\frac{k^2}{4}}{t(t-1)(t+1)}\psi_k =0\ .
\end{equation}
This is a general Heun equation with parameters
\begin{equation}
a=-1\,, \qquad q=-\frac{k^2}{4}\,, \qquad \alpha=\beta=\frac34\,, \qquad \gamma=\frac12\,, \qquad \delta=1\,,
\end{equation}
in the standard Heun notation. A convenient basis of solutions near the conformal boundary is
\begin{align}
E_k(z) &\equiv \operatorname{HeunG}\!\left(-1,-\frac{k^2}{4};\frac34,\frac34,\frac12,1;z^2\right)\,, \\
O_k(z) &\equiv z\operatorname{HeunG}\!\left(-1,-\frac{k^2}{4};\frac54,\frac54,\frac32,1;z^2\right)\,,
\end{align}
where $\operatorname{HeunG}$ denotes the Heun function. The behavior of these functions near $z=0$ reads
\begin{equation}\label{properties-Ek-Ok}
E_k(0)=1\,, \qquad E_k'(0)=0\,, \qquad O_k(0)=0\,, \qquad O_k'(0)=1\ .
\end{equation}
On the other hand, the solution regular at the horizon and normalized to unity is
\begin{equation}\label{eq:horizon-regular-heun}
R_k(z) = \operatorname{HeunG}\!\left(2,\frac{k^2}{4}+\frac{9}{16};\frac34,\frac34,1,\frac12;1-z^2\right)\ .
\end{equation}
It satisfies
\begin{equation}\label{horizon-regularity-R-Rprime}
   R_k(1)=1\,, \qquad R_k'(1) = -\frac{k^2+\frac94}{4}\ .
\end{equation}
The solution $R_k(z)$ may be written as a linear combination of $E_k(z)$ and $O_k(z)$,
\begin{equation}
    R_k(z)=A(k)E_k(z)+B(k)O_k(z)\ .
\end{equation}
Thanks to properties \eqref{properties-Ek-Ok}, the coefficients $A(k)$, $B(k)$ may be obtained by expanding $R_k(z)$ near the conformal boundary $z=0$,
\begin{equation}
   R_k(z) = A(k)+B(k)z+O(z^2)\,,
\end{equation}
i.e. $A(k)=R_k(0)$ and $B(k)=R_k'(0)$. The solution $\psi_k(z)$ that we are after is proportional to $R_k(z)$, with the overall normalization fixed by the condition \eqref{eq:boundary-derivative-condition},
\begin{equation}\label{eq:normalized-radial-solution}
\psi_k(z) = -4\pi\,\frac{R_k(z)}{B(k)}\ .
\end{equation}
It follows from \eqref{gtilde-psi} that the exact momentum-space correlator is
\begin{equation}\label{eq:exact-momentum-space-correlator}
\widetilde{g}^{(0)}(k) = -4\pi\,\frac{A(k)}{B(k)} = -4\pi\,\frac{R_k(0)}{R_k'(0)}\ .
\end{equation}

\subsection{Position-space representation}

By performing the inverse Fourier transform for a rotationally invariant function in three spatial dimensions, we can write the position space function $g^{(0)}(x)$ as
\begin{equation}\label{eq:exact-position-space-correlator}
g^{(0)}(x) = \frac{1}{2\pi^2x} \int_0^\infty dk\, k\sin(kx)\, \widetilde{g}^{(0)}(k) = -\frac{2}{\pi x} \int_0^\infty dk\, k\sin(kx)\, \frac{A(k)}{B(k)}\,,
\end{equation}
where in the second step we used \eqref{eq:exact-momentum-space-correlator}. In evaluating the last integral, it is useful to isolate the pure-AdS singularity. Since the pure-AdS solution corresponds to
\begin{equation}\label{AB AdS}
\frac{A_{\mathrm{AdS}}(k)}{B_{\mathrm{AdS}}(k)} = -\frac{1}{k}\,,
\end{equation}
the finite part of the correlator can be written as
\begin{equation}\label{eq:finite-exact-position-space-correlator}
g^{(0)}(x)-\frac{2}{\pi x^2} = -\frac{2}{\pi x} \int_0^\infty dk\, \sin(kx) \left[1+k\frac{R_k(0)}{R_k'(0)}\right]\ .
\end{equation}
The quantity in square brackets decays as $k^{-4}$ at large $k$, so the integral is convergent.

While the expressions \eqref{eq:exact-momentum-space-correlator} and \eqref{eq:finite-exact-position-space-correlator} are exact, extracting from them expansion coefficients of $g^{(0)}(x)$ requires some work. We distinguish between two types of contributions. Firstly, the function \eqref{eq:exact-momentum-space-correlator} admits an asymptotic expansion at large $k$ (see details in Appendix \ref{app:Near-boundary expansion}). To a few lowest orders,
\begin{equation}\label{asymptotic-expansion-tildeg}
\widetilde{g}^{(0)}(k) = \frac{4\pi}{k} +\frac{3\pi}{4k^5} +\frac{2637\pi}{128k^9} +O\left(k^{-13}\right)\ .
\end{equation}
Using the Fourier-transform identities (we write $\mathcal{F}^{d=3}$ to denote the three-dimensional Fourier transform)
\begin{equation}
\mathcal{F}^{d=3}\left[x^2\log x\right] = \frac{12\pi^2}{k^5}\,, \qquad 
\mathcal{F}^{d=3}\left[x^6\log x\right] = \frac{10080\pi^2}{k^9}\,,
\end{equation}
one obtains the short-distance expansion
\begin{equation}\label{eq:exact-short-distance-expansion}
g^{(0)}(x)-\frac{2}{\pi x^2} = \gamma_0 + \frac{x^2}{16\pi}\log x + \gamma_1 x^2 + \gamma_2 x^4 + \frac{293}{143360\pi}x^6\log x
+\cdots\ .
\end{equation}
Coefficients $\gamma_i$ of non-logarithmic terms in \eqref{eq:exact-short-distance-expansion} are not captured by the asymptotic expansion \eqref{asymptotic-expansion-tildeg}. Rather, they can be obtained by integrating over the exact expression \eqref{eq:exact-momentum-space-correlator}. In practice, the integrals are done numerically. For instance, the leading coefficient $\gamma_0$ reads
\begin{equation}\label{gamma0}
    \gamma_0 = -\frac{2}{\pi} \int_0^\infty dk\, k
    \left(
    1 + k\,\frac{A(k)}{B(k)}
    \right) \approx -0.02357608208468\ .
\end{equation}
An efficient way to evaluate this integral is to integrate numerically up to a finite cutoff $\Lambda$ and use the asymptotic expansion of the integrand to account for the remaining contribution from $k>\Lambda$. Explicit calculations of the lowest coefficients $\gamma_1$ and $\gamma_2$ are presented in Appendix \ref{app:Higher expansion coefficients}. Values of some further low-lying coefficients are collected in Table \ref{tab:zero-mode-coefficients}.

\subsection{Results}

In this subsection, we show the results obtained by the method described above. Let us give an executive summary of the process. By solving the PDE in the bulk, we get expansion coefficients $\gamma_k$ of the zero mode listed in Table \ref{tab:zero-mode-coefficients}. These are substituted in equation \eqref{extracting-expansion-coefficients-simpler} to obtain coefficients $a_{k,0}$. Let us note that the infinite sum in the first line of \eqref{extracting-expansion-coefficients-simpler} diverges and is resummed using Borel resummation. Coefficients $a_{k,i}$ with $i>0$ are obtained in equations \eqref{coefficients-aki-i>0} and \eqref{coefficients-aki-Borel}. Finally, the double-trace coefficients $b_{m,J}$ are given in terms of $a_{k,i}$ by the linear transformation \eqref{double-trace-from-expansion-Section2}, with the transformation matrix \eqref{double-trace-from-expansion-matrix}.
\begin{table}[t]
    \centering
    \renewcommand{\arraystretch}{1.9}
    \begin{tabular}{|c|c|}
        \hline
        $k$
        &
        $\left. g^{(0)}(x) \right|_{x^{2k}} = \gamma_k + \tilde\gamma_k \log x
        $
        \\
        \hline
        0 & $-0.02357608208468$ \\
        1 & $-0.01617472265814 + \displaystyle\frac{1}{16\pi}\log x$ \\
        2 & $0.0003249565648$ \\
        3 & $-0.000624722994
             + \frac{293}{143360\pi}\log x$ \\
        4 & $-1.90983\cdot 10^{-6}$ \\
        \hline
    \end{tabular}
    \caption{Coefficients of the terms proportional to $x^{2k}$ in the
    small-$x$ expansion of the zero mode $g^{(0)}(x)$.}
    \label{tab:zero-mode-coefficients}
\end{table}

\paragraph{Example: computation of $b_{0,0}$.}
Let us illustrate the procedure for the lowest double-trace
coefficient $b_{0,0}$. Setting $k=0$ in
\eqref{extracting-expansion-coefficients-simpler}, we obtain
\begin{equation}
    a_{0,0} = \gamma_0 +
    2\sum_{i=0}^{\infty}
    a^{i}_{0,2i}\,
    \zeta(3-4i)\ .
\end{equation}
The series is asymptotic and is evaluated using the Borel-resummation
prescription described above. Equivalently,
\begin{equation}
    a_{0,0} = \gamma_0 + 2\int_0^\infty dt\,e^{-t} \sum_{i=0}^\infty \frac{a^i_{0,2i}}{(4i)!} \zeta(3-4i)t^{4i}\ .
\end{equation}
Substituting\footnote{The factor of $\pi^3$ arises from rescaling the inverse temperature from $\beta=\pi$, as in \eqref{gamma0}, to $\beta=1$.}
\begin{equation}
    \gamma_0=-0.02357608208468\,\pi^3\,,
\end{equation}
and evaluating the Borel-resummed image contribution gives
\begin{equation}
    a_{0,0} = 1.113079193089\ .
\end{equation}
For $m=J=0$, the change-of-basis formula
\eqref{double-trace-from-expansion-Section2} reduces to
\begin{equation}
    b_{0,0}  = a_{0,0}\mathcal{M}^{(0,0)}_0 = a_{0,0}\ .
\end{equation}

\paragraph{Example: computation of $b_{1,0}$ and $b_{0,2}$.} Let us further illustrate the procedure by computing the two double-trace coefficients $b_{1,0}$ and $b_{0,2}$. We first need the coefficients $a_{1,0}$ and $a_{0,1}$ of the monomial basis. Setting $k=1$ in \eqref{extracting-expansion-coefficients-simpler}, we obtain
\begin{align}
a_{1,0}
={}& \gamma_1 +2\sum_{j=0}^1
\sum_{\substack{i=j\\ i\neq1}}^\infty
a^i_{j,2i-j}\binom{2i-j-\frac32}{1-j}\zeta(5-4i)\nonumber\\
&+
\sum_{j=0}^1 a^1_{j,2-j}\binom{\frac12-j}{1-j}
\left[-2\log 2+\gamma_{\rm E}-\psi(2-j)\right]\ .
\end{align}
The infinite series is asymptotic and is evaluated using the Borel-resummation prescription described above. That is,
\begin{align}
a_{1,0}
={}&
\gamma_1
+2\int_0^\infty dt\,e^{-t}
\Bigg[
\sum_{\substack{i=0\\ i\neq1}}^\infty
\frac{a^i_{0,2i}}{(4i)!}
\left(2i-\frac32\right)
\zeta(5-4i)t^{4i}
+
\sum_{i=2}^\infty
\frac{a^i_{1,2i-1}}{(4i)!}
\zeta(5-4i)t^{4i}
\Bigg]
\nonumber\\
&+
\frac{a^1_{0,2}}{2}
\left(-2\log 2+2\gamma_{\rm E}-1\right)
+
a^1_{1,1}
\left(-2\log 2+2\gamma_{\rm E}\right)\ .
\end{align}
Furthermore, after rescaling the inverse temperature from $\beta=\pi$
to $\beta=1$, the zero-mode contribution is
\begin{equation}
\gamma_1 = -0.01617472265814\,\pi^5
+\frac{\pi^4}{16}\log\pi\ .
\end{equation}
Evaluating the Borel-resummed image contribution gives
\begin{equation}
2\int_0^\infty dt\,e^{-t}
\Bigg[
\sum_{\substack{i=0\\ i\neq1}}^\infty
\frac{a^i_{0,2i}}{(4i)!}
\left(2i-\frac32\right)
\zeta(5-4i)t^{4i}
+
\sum_{i=2}^\infty
\frac{a^i_{1,2i-1}}{(4i)!}
\zeta(5-4i)t^{4i}
\Bigg]
=
-2.8086515\ .
\end{equation}
Putting everything together, we find
\begin{equation}
    a_{1,0} = -1.909862909813\ .
\end{equation}
The second monomial coefficient follows from
\eqref{coefficients-aki-i>0}. For $k=0$ and $l=1$, we obtain
\begin{equation}
a_{0,1} = 2\sum_{i=0}^\infty
a^i_{0,2i}
\binom{4i-3}{2}
\zeta(5-4i)\ .
\end{equation}
Again, the series is asymptotic and is evaluated by Borel resummation:
\begin{equation}
a_{0,1} = 2\int_0^\infty dt\,e^{-t}
\sum_{i=0}^\infty
\frac{a^i_{0,2i}}{(4i)!}
\binom{4i-3}{2}
\zeta(5-4i)t^{4i} =
7.686391301178\ .
\end{equation}
Finally, using the change-of-basis formula \eqref{double-trace-from-expansion-Section2}, we get
\begin{align*}
b_{1,0} & = a_{0,1}\mathcal{M}^{(1,0)}_0 + a_{1,0}\mathcal{M}^{(1,0)}_1 = 
\frac14 a_{0,1} + \frac34 a_{1,0}=0.48920064\,,\\
b_{0,2} & =
a_{0,1}\mathcal{M}^{(0,2)}_0 + a_{1,0}\mathcal{M}^{(0,2)}_1
= \frac14 a_{0,1} - \frac14 a_{1,0}= 2.39906355\ .
\end{align*}
Similarly, we obtain the following double-trace thermal coefficients 
\begin{equation}\label{main-numbers}
\begin{array}{ccccccc}
&&&
\begin{gathered}
b_{0,0}\\[-0.2em]
1.113079193089
\end{gathered}
&&&
\\[0.8em]
&&
\begin{gathered}
b_{1,0}\\[-0.2em]
0.4892006
\end{gathered}
&&
\begin{gathered}
b_{0,2}\\[-0.2em]
2.3990636
\end{gathered}
&&
\\[0.8em]
&
\begin{gathered}
b_{2,0}\\[-0.2em]
1.42104
\end{gathered}
&&
\begin{gathered}
b_{1,2}\\[-0.2em]
0.86546
\end{gathered}
&&
\begin{gathered}
b_{0,4}\\[-0.2em]
6.853229
\end{gathered}
&
\\[0.8em]
\begin{gathered}
b_{3,0}\\[-0.2em]
54.4327
\end{gathered}
&&
\begin{gathered}
b_{2,2}\\[-0.2em]
21.46210
\end{gathered}
&&
\begin{gathered}
b_{1,4}\\[-0.2em]
-37.7555
\end{gathered}
&&
\begin{gathered}
b_{0,6}\\[-0.2em]
20.38886
\end{gathered}
\end{array}
\end{equation}
The coefficient $b_{0,0}$ was obtained in several works, \cite{Parisini:2023nbd,Buric:2025fye,Niarchos:2026wsw}, and our value is consistent with those. Furthermore, the values of $b_{1,0}$ and $b_{0,2}$ are consistent with the recent results of \cite{Niarchos:2026wsw}. For all these coefficients, we are able to determine many more digits than were previously available. The other double-trace coefficients in \eqref{main-numbers} are, to the best of our knowledge, new.

\section{Analytic structure in real time}
\label{S:Analytic structure in real time}

In this short section, we discuss analytic continuation of the two-point function to complex time in order to address the following questions:
\begin{enumerate}
    \item Does the Bessel representation \eqref{g-images-K}\footnote{Throughout this section, whenever referring to representation \eqref{g-images-K}, what we mean is its Borel resummation.} admit a well-defined continuation to real time?
    \item Do the complex bulk-cone singularity curves found in the stress-tensor sector of the correlator in \cite{Araya:2026shz} persist in the full two-point function?
\end{enumerate}

The first will be answered in the positive and the second in the negative.

\subsection{Analytic continuation to real time}

\begin{figure}[t]
    \centering
    \includegraphics[width=0.85\textwidth]
    {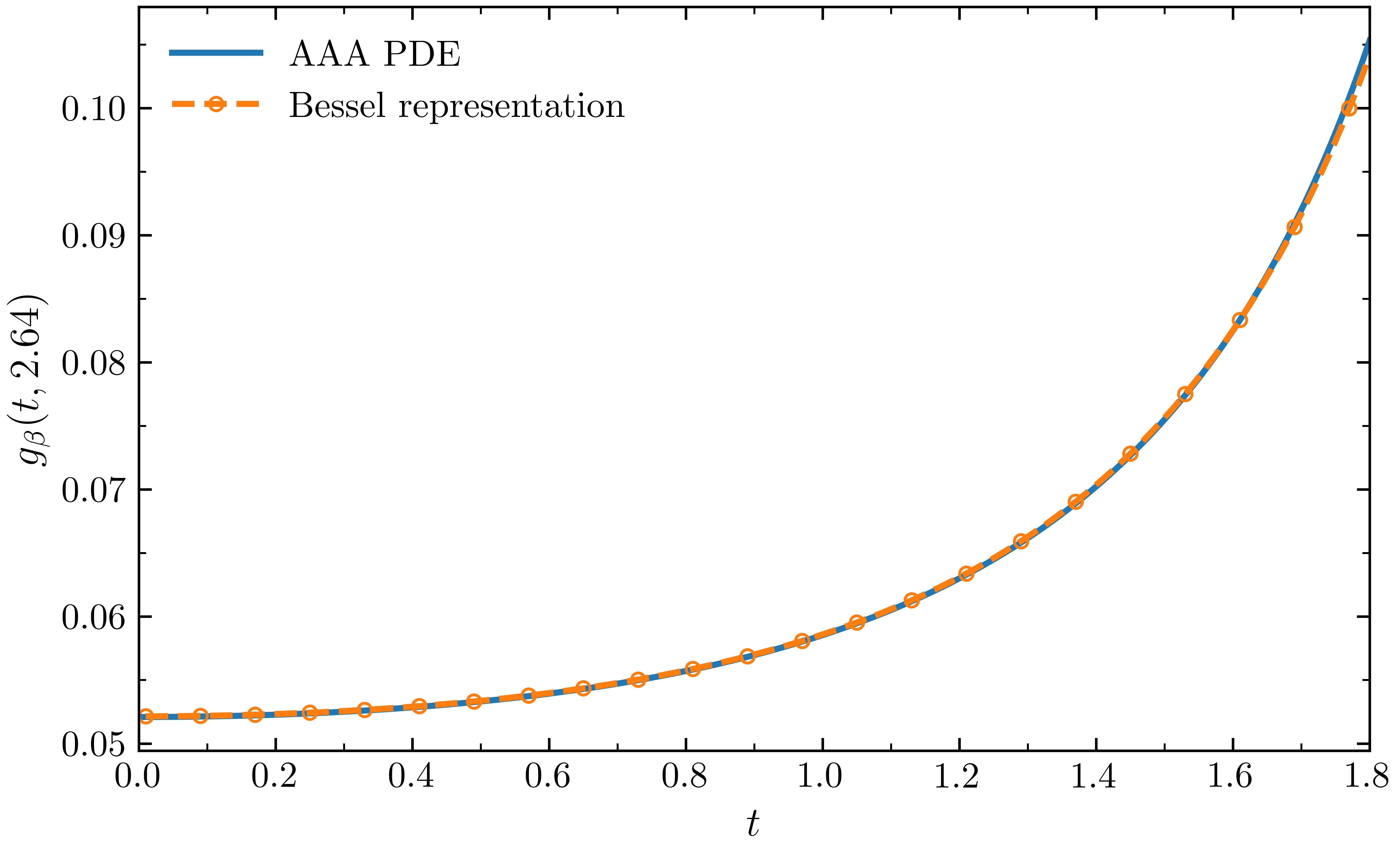}
    \caption{Comparison of the numerical AAA PDE solution and the Bessel representation for $g_\beta(t,2.64)$. The two-point function exhibits only the standard light-cone singularity, located at $t=x\approx 2.64$. We set $\beta = \pi$.}
    \label{fig:comparison}
\end{figure}

To address the first question, we fix $x>0$ and regard the two-point function as a function of $t=-i\tau$. Then it is sufficient to consider only the contribution from the nonzero modes, given in the second line of \eqref{g-images-K}. As discussed in Appendix \ref{app:gff_bessel}, we expect the two-point function representation to be valid for $|t|=|\operatorname{Im}\tau| < x$. This is the region for which the Bessel representation \eqref{Bessel GFF} of the GFF two-point function holds. However, since passing from \eqref{Bessel GFF} to \eqref{g-images-K} involves an infinite sum, we further verify the validity of \eqref{g-images-K} in the region $|t|=|\operatorname{Im}\tau| < x$ by comparing it to the analytic continuation of the solution obtained from the (three-variable) bulk PDE. The bulk PDE and its numerical solution have been extensively discussed in \cite{Parisini:2023nbd,Buric:2025fye}. We start with the numerical solution obtained in \cite{Buric:2025fye}, which applies for real $\tau$. For each fixed value of $x$, we evaluate this solution at a set of real Euclidean times $\{\tau_\ell\}$ and apply the adaptive Antoulas--Anderson (AAA) algorithm \cite{AntoulasAnderson1986,Nakatsukasa_2018} to these samples. AAA selects support points $z_j$ adaptively from the sampled $\tau$-grid and constructs a rational approximant in barycentric form,
\begin{equation}\label{AAA-approximant}
    R_x(\tau)=
    \frac{\displaystyle\sum_{j=1}^{M}
    \frac{w_j\,g_\beta(z_j,x)}{\tau-z_j}}
    {\displaystyle\sum_{j=1}^{M}
    \frac{w_j}{\tau-z_j}}\ .
\end{equation}
At each step, the next support point is chosen where the current approximation has its largest residual on the Euclidean data, while the weights $w_j$ are determined from a linear least-squares problem. Thus, the algorithm concentrates support points in the parts of the Euclidean interval that are hardest to approximate, without requiring us to prescribe the poles of $R_x$ in advance. Numerical noise may generate Froissart doublets, namely nearly coincident pole-zero pairs with negligible effect on the approximated Euclidean data. We remove these spurious pairs before continuing away from the real axis. The cleaned approximant is then evaluated at $\tau=it$, giving
\begin{equation}
    g_\beta(t,x)\big|_{\mathrm{PDE/AAA}}=R_x(it)\ .
\end{equation}
AAA is used here only to analytically continue the independent numerical bulk solution; it does not enter the construction or Borel resummation of the Bessel representation. Figure \ref{fig:comparison} compares the Bessel result with this numerical PDE/AAA continuation in the domain $0<t<x$ where both procedures are expected to apply. Their very good agreement provides further support for the validity of \eqref{g-images-K}.

\subsection{No complex bulkcone singularities}

\begin{figure}[t]
    \centering
    \includegraphics[width=0.8\textwidth]
    {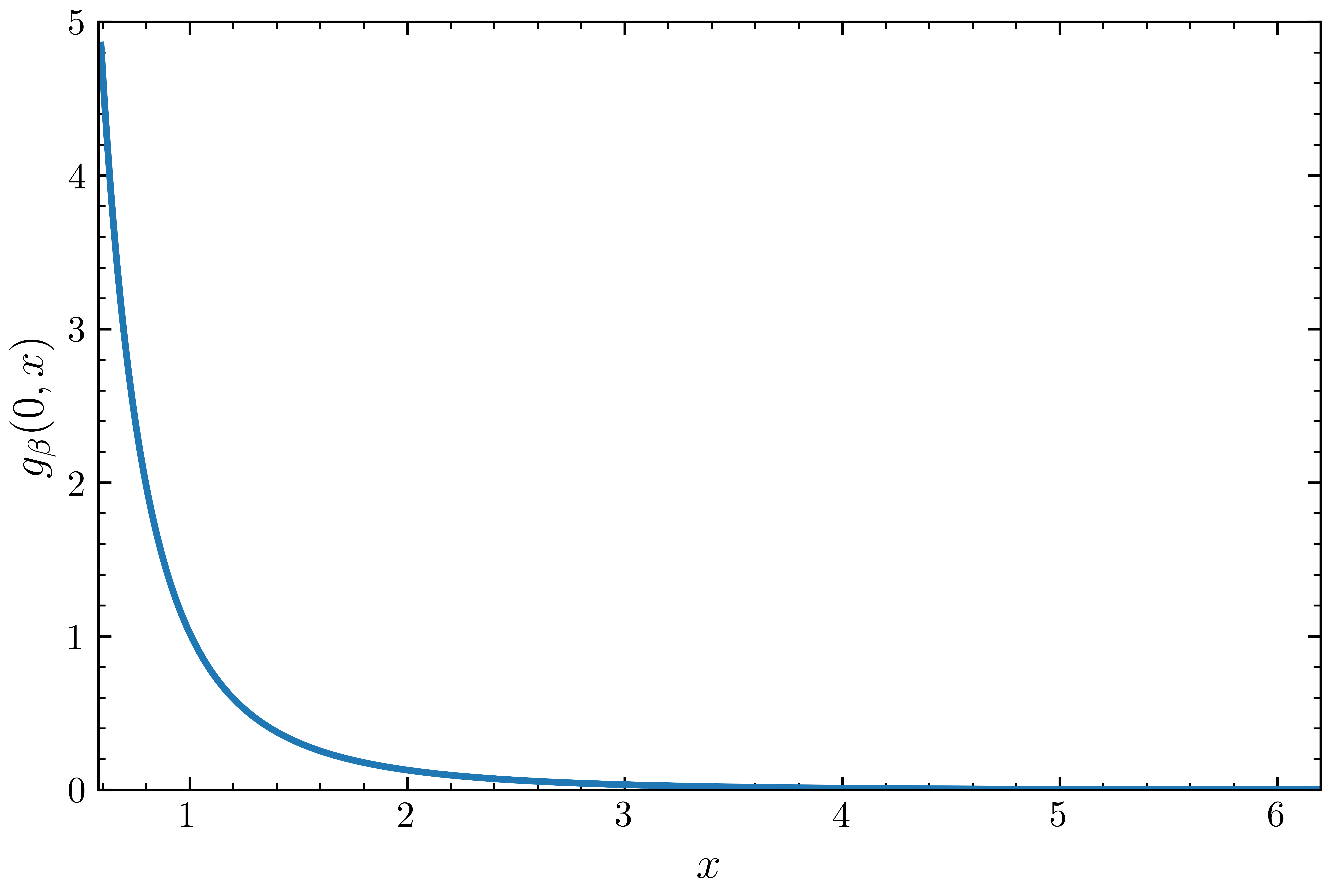}
    \caption{Numerical PDE solution for $g_\beta(0,x)$. The two-point function exhibits only the standard light-cone singularity, located at $x=t=0$. We set $\beta = \pi$.}
    \label{fig:t=0}
\end{figure}

As shown in \cite{Araya:2026shz}, the stress-tensor sector $g^T_\beta(t,x)$ of the thermal two-point function exhibits singularities for all $x\geq x_{\mathrm{c}}$, where
\begin{equation}\label{x_c}
x_{\mathrm{c}} \equiv \frac{\sqrt{\pi}}{2}\frac{\Gamma(1/4)}{\Gamma(3/4)} \approx 2.622\,, \qquad t_{\mathrm{c}} = \pm\frac{\pi}{2}\ .
\end{equation}
We choose $x_* = 2.64$, slightly above $x_{\mathrm{c}}$. At this value, \cite{Araya:2026shz} predicts two pairs of bouncing singularities in the stress-tensor sector, symmetrically located about $t=\pm \frac{\pi}{2}$, namely
\begin{equation}
t=\pm\left(\frac{\pi}{2}\pm 0.21752\right) \approx \pm 1.35328\,,\ \pm 1.78832\ .
\end{equation}
As shown in Figure \ref{fig:comparison}, neither of the methods we described predicts a singularity of the Wightman two-point function at these locations. We therefore conclude that the complex bulk-cone singularities present in the stress-tensor sector of the correlator do not persist in the full two-point function. To further corroborate this conclusion, we consider a second example. Setting $t=0$, we consider the correlator as a function of $x$. According to \cite{Araya:2026shz}, the stress-tensor sector is expected to develop a singularity near $x\approx 3.37$. In this case, no AAA-based analytic continuation to real time is required: the PDE solution at $\tau=0$ is sufficient to probe the presence of poles along the real $x$-axis. Figure \ref{fig:t=0} shows that the two-point function remains regular for all $x>0$.

\section{Discussion}
\label{S:Discussion}

In this paper, we extended the method of \cite{Buric:2025fye} for computing double-trace thermal coefficients in holography to obtain these coefficients refined by spin. Our approach takes as the starting point the stress-tensor sector of the thermal two-point function and performs its sum over images using analytic continuation. Finally, the zero Fourier mode of the two-point function, which is left undetermined by this process, is fixed by solving a two-variable PDE in the bulk. The latter reduces to an exactly solvable Heun equation in momentum space. While the construction applies for a field of general conformal dimension, in our numerical studies we focused on $\Delta_\phi=3/2$. The main results are the low-lying double-trace thermal coefficients given in equation \eqref{main-numbers}. For the coefficients that have been computed in the literature, notably in \cite{Niarchos:2026wsw}, we observe complete agreement. Our results also include new coefficients that have not been obtained previously.
\smallskip

An interesting extension of our results would be to repeat the analysis in finite volume, i.e. for a CFT on $S^1 \times S^3$, by working with the non-planar AdS-Schwarzschild black hole. The CFT data probed by correlators on $S^1\times S^{d-1}$ are the flat-space OPE coefficients, \cite{Gobeil:2018fzy,Buric:2024kxo}, thus being of interest in various well-established contexts. Other, more direct, generalizations of this work include deriving double-trace thermal coefficients for generic values of $\Delta_\phi$ and in other spacetime dimensions (with even-dimensional boundary). More generally, the resolution of CFT data in spin allows us to study various scaling limits, from the lightcone regime $(\Delta\to \infty,\Delta-J=\text{finite})$, \cite{Fitzpatrick:2012yx,Komargodski:2012ek}, to the ‘thermal EFT' regime $(\Delta\to\infty,J=\text{finite})$, \cite{Benjamin:2023qsc,Benjamin:2024kdg,Buric:2025uqt,Buric:2026pes}. In both of these limits, CFT data exhibits universal properties, being interpolated by a semi-universal behavior in between, \cite{Anand:2025mfh,Komargodski:2026ain}. It would be interesting to explore how the thermal coefficients behave in these various regimes. 
\smallskip

An additional question, relating the thermal two-point function to spacetime geometry, is whether the complex-geodesic scale $x_{\mathrm c}$ given in \eqref{x_c} is encoded in the asymptotic distribution of poles in complex spatial momentum. At zero frequency, setting $k=i\kappa$ turns the radial equation \eqref{eq:radial-zero-frequency} into a Sturm--Liouville problem. Its leading WKB quantization gives
\begin{equation}
    \kappa_{n+1}-\kappa_n \longrightarrow \frac{\pi}{L_{\rm opt}}\,, \qquad
    L_{\rm opt}\equiv\int_0^1\frac{dz}{\sqrt{1-z^4}} =\frac{x_{\mathrm c}}{2}\,,
\end{equation}
where $\kappa_n$ are the eigenvalues, or equivalently, the poles of $\tilde g$. Consequently, the asymptotic pole density satisfies
\begin{equation}
    \frac{dn}{d\kappa} \longrightarrow \frac{x_{\mathrm c}}{2\pi}\,, \qquad n\to\infty\ .
\end{equation}
The leading density is independent of a fixed scalar mass, while standard versus alternative quantization affects the constant offset in the asymptotic pole sequence. The exact zero-mode Heun solution obtained in this work therefore provides a direct way to test this prediction by locating its high complex-$k$ poles. The Bohr--Sommerfeld condition of \cite{Festuccia:2008zx} suggests a possible extension to nonzero frequency. Formally continuing that condition to complex $k$ at fixed $\omega$ leads to a density controlled by the derivative of the WKB action. Its validity, however, requires verifying that the relevant turning points, branches, and anti-Stokes contour continue to the complex-$k$ region under consideration.
\smallskip

It is worth noting that the same radial optical length $L_{\rm opt}$ appears in the holographic glueball problem of \cite{Csaki:1998qr,Minahan:1998tm}. Under double Wick rotation, the black-brane geometry becomes the AdS soliton, one of the simplest holographic models of confinement. Normalisability at the AdS boundary is preserved, while regularity at the static black-brane horizon maps to regularity at the smooth tip of the soliton. Therefore, for the same bulk scalar and the same boundary quantization, the two radial boundary-value problems are identical, and soliton normal-mode masses and the black-brane (screening) eigenvalues coincide, $M_n^{(\Delta_\phi)}=\kappa_n^{(\Delta_\phi)}$. In particular, the massless-dilaton glueball spectrum of \cite{Csaki:1998qr,Minahan:1998tm} coincides with the static screening spectrum of the corresponding $\Delta_\phi=4$ scalar. Although the explicit poles studied here apply instead to $\Delta_\phi=3/2$, the bulk mass and boundary quantization modify only the subleading WKB phase. Hence, for both values of $\Delta_\phi$, we have the universal large-radial-excitation spacing
\begin{equation}
    M_{n+1}-M_n\longrightarrow \frac{\pi}{L_{\rm opt}}\ .
\end{equation}
This provides another physical interpretation of $L_{\rm opt}$ as the radial length controlling the asymptotic spectral density of both screening states and soliton glueballs.
\smallskip

Finally, our results provide an example of an explicit formula for holographic CFT data featuring Heun connection coefficients. In the case at hand, the connection problem was sufficiently simple and its solution did not require extensive theory. It would be interesting to explore what other CFT data have equally explicit representations in terms of Heun functions and whether recent developments on the connection problem, see \cite{Bonelli:2022ten} and references therein, can provide efficient computational tools for holographic CFTs.

\section*{Acknowledgements}

We wish to thank J. Barrat, D. N. Bozkurt, M. Dodelson, C. Esper, M. Kulaxizi, E. Marchetto, A. Miscioscia, V. Niarchos, E. Pomoni and S. Valach for discussions. This publication has emanated from research conducted with the financial support of Taighde Éireann – Research Ireland under Grant number SFI-22/FFP-P/11444.

\appendix

\section{Change of basis for thermal conformal blocks}
\label{app:change-of-basis}

In this appendix, we collect the change-of-basis formulae for double-trace and stress-tensor sector expansions. Throughout this appendix, we set $\beta=1$.

\subsection{Double-trace sector}
\label{app:double-trace-change-of-basis}

We now derive the change of basis from the coefficients $a_{k,i}$ of the monomial expansion in $x^{2}$ and $\tau^{2}$ to the double-trace coefficients $b_{m,J}$. The monomial expansion is
\begin{equation}
    g_\beta^{\rm DT}(\tau,x) = \sum_{k=0}^{\infty} \sum_{i=0}^{\infty} a_{k,i}\,x^{2k}\tau^{2i} =
    \sum_{q=0}^{\infty} \sum_{k=0}^{q}
    a_{k,q-k}\, x^{2k}\tau^{2(q-k)}\ .
    \label{dt-monomial-expansion}
\end{equation}
In the second equality, we have grouped terms by the total degree $q=k+i$. Using relations \eqref{change-of-vars-1}-\eqref{change-of-vars-2}, we can write
\begin{equation}
    g_\beta^{\rm DT}(\tau,x)
    =
    \sum_{q=0}^{\infty}
    r^{2q}
    \sum_{k=0}^{q}
    a_{k,q-k}
    (1-\eta^2)^k
    \eta^{2(q-k)} .
    \label{dt-monomial-eta-expansion}
\end{equation}
On the other hand, the conformal block decomposition of the double-trace
sector is
\begin{equation}
    g_\beta^{\rm DT}(\tau,x) = \sum_{m=0}^\infty \sum_{J=0,2,\ldots}^{\infty} b_{m,J}\,r^{2m+J}C_J^{(1)}(\eta)\ .
    \label{dt-block-expansion}
\end{equation}
It is useful to group the expansion by the total radial order $q=m+J/2$. Then
\begin{equation}
    g_\beta^{\rm DT}(\tau,x) = \sum_{q=0}^\infty r^{2q} \sum_{J=0,2,\ldots,2q} b_{q-J/2,J}\, C_{J}^{(1)}(\eta)\ .
    \label{dt-fixed-q-expansion}
\end{equation}
Equating this expression with \eqref{dt-monomial-eta-expansion} order by order in $r$, we find
\begin{equation}\label{dt-fixed-q-basis-relation}
    \sum_{J=0,2,\ldots,2q} b_{q-J/2,J}\, U_J(\eta) = \sum_{k=0}^q a_{k,q-k} (1-\eta^2)^k \eta^{2(q-k)}\ .
\end{equation}
Here, $U_J(\eta)$ are Chebyshev polynomials of the second kind,
\begin{equation}
    C_J^{(1)}(\eta) = U_J(\eta)\,, \qquad U_J(\cos\theta) = \frac{\sin((J+1)\theta)}{\sin\theta}\ .
\end{equation}
They satisfy orthogonality relations
\begin{equation}
    \int_{-1}^1 d\eta\, \sqrt{1-\eta^2}\, U_J(\eta)U_{J'}(\eta) = \frac{\pi}{2}\delta_{J,J'}\ .
\end{equation}
We use these relations to project both sides of \eqref{dt-fixed-q-basis-relation} onto the Chebyshev basis, obtaining
\begin{equation}
    b_{q-J/2,J} = \frac{2}{\pi} \int_{-1}^1 d\eta\, \sqrt{1-\eta^2}\,
    U_J(\eta) \sum_{k=0}^q a_{k,q-k} (1-\eta^2)^k \eta^{2(q-k)}\ .
\end{equation}
Equivalently, setting $q=m+\frac{J}{2}$, we get
\begin{equation}\label{dt-change-of-basis-final}
    b_{m,J} = \sum_{k=0}^{m+J/2} a_{k,m+J/2-k} \mathcal M^{(m,J)}_k\,,
\end{equation}
where
\begin{equation}\label{dt-change-of-basis-integral}
    \mathcal M^{(m,J)}_k =
    \frac{2}{\pi}\int_{-1}^{1} d\eta\, \sqrt{1-\eta^2}\, U_J(\eta)(1-\eta^2)^k \eta^{2(m+J/2-k)}\ .
\end{equation}
For even spin $J$, we use the explicit expansion
\begin{equation}
    U_J(\eta) = \sum_{r=0}^{J/2} (-1)^r 2^{J-2r} \binom{J-r}{r} \eta^{J-2r}\ .
\end{equation}
Substituting this into \eqref{dt-change-of-basis-integral} gives
\begin{equation}\label{dt-change-of-basis-matrix}
    \mathcal M^{(m,J)}_k =
    \frac{2}{\pi} \sum_{r=0}^{J/2} (-1)^r
    2^{J-2r} \binom{J-r}{r} B\!\left( m+J-k-r+\frac12,\,k+\frac32 \right)\ .
\end{equation}

\subsection{Stress-tensor sector}
\label{app:stress-tensor-change-of-basis}

The conformal block decomposition of the stress-tensor sector reads
\begin{equation} g_\beta^T(\tau,x) = \frac{1}{r^{2\Delta_\phi}} \sum_{n=0}^{\infty} \sum_{J=0,2,\ldots,2n} \lambda_{n,J}\, r^{4n} C_J^{(1)}(\eta)\,, \label{app:gT-block-expansion} \end{equation}
while the near-boundary expansion is given by
\begin{equation}
    g_\beta^T(\tau,x) = \frac{1}{r^{2\Delta_\phi}}\sum_{i=0}^\infty
    r^{4i} \sum_{j=0}^i a^i_{j,2i-j}(1-\eta^2)^j\ .
\end{equation}
Equating this expression with \eqref{app:gT-block-expansion} order by order in $r$, we find
\begin{equation}\label{app:stress-basis-equation}
    \sum_{J=0,2,\ldots,2n} \lambda_{n,J} C_J^{(1)}(\eta) = \sum_{j=0}^n a^n_{j,2n-j}(1-\eta^2)^j\ .
\end{equation}
Projecting \eqref{app:stress-basis-equation} onto $U_J(\eta)$ gives
\begin{align}
    \lambda_{n,J} = \frac{2}{\pi}\int_{-1}^1
    d\eta\,\sqrt{1-\eta^2}\,U_J(\eta)&\sum_{j=0}^n a^n_{j,2n-j}(1-\eta^2)^j\\
    = &\sum_{j=0}^n a^n_{j,2n-j}\,
    \frac{(-1)^{J/2}4^{-j}\Gamma(2j+2)}{\Gamma(j+1-J/2)\Gamma(j+2+J/2)}\,, \nonumber
\end{align}
where $J=0,2,\dots,2n$. In summary, defining the change-of-basis matrix
\begin{equation}
    \mathcal{M}_{jJ}
    \equiv\frac{(-1)^{J/2}4^{-j}\Gamma(2j+2)}{\Gamma(j+1-J/2)\Gamma(j+2+J/2)}\,,
\end{equation}
we may write
\begin{equation}\label{app:stress-change-of-basis}
    \lambda_{n,J} = \sum_{j=0}^n \mathcal{M}_{jJ}\,a^n_{j,2n-j}\,, \qquad J=0,2,\ldots,2n\ .
\end{equation}

\section{Bessel representation of the GFF image sum}
\label{app:gff_bessel}

In this appendix, we derive the Bessel-function representation \eqref{Bessel GFF} used in the main text. We start from the GFF image sum
\begin{equation}
g^{\rm GFF}_{\Delta}(\tau,x) = \sum_{m\in\mathbb{Z}} \frac{1}{\big[(\tau+m)^2+x^2\big]^{\Delta}}\ .
\end{equation}
For the moment, we assume $\Delta>1/2$, so that the image sum is absolutely convergent for $x>0$. The final expression will then provide an analytic continuation in $\Delta$. Let
\begin{equation}
f(u)=\frac{1} {(u^2+x^2)^{\Delta}}\ .
\end{equation}
Then
\begin{equation}
g^{\rm GFF}_{\Delta}(\tau,x) = \sum_{m\in\mathbb Z} f(m+\tau)\ .
\end{equation}
Applying Poisson summation gives
\begin{equation}
\sum_{m\in\mathbb Z} f(m+\tau) = \sum_{n\in\mathbb Z} e^{2\pi i n\tau}\widehat f(n)\,,
\end{equation}
where
\begin{equation}
\widehat f(n) = \int_{-\infty}^{\infty}du\, \frac{e^{-2\pi i n u}}{(u^2+x^2)^{\Delta}}\ .
\end{equation}
It remains to compute this Fourier transform. We use the Schwinger representation
\begin{equation}
\frac{1}{(u^2+x^2)^{\Delta}} = \frac{1}{\Gamma(\Delta)}
\int_0^\infty ds\, s^{\Delta-1}e^{-s(u^2+x^2)}\ .
\end{equation}
Substituting this into the Fourier transform, one obtains
\begin{align}
\widehat f(n) &= \frac{1}{\Gamma(\Delta)} \int_0^\infty ds\,s^{\Delta-1}e^{-s x^2} \int_{-\infty}^{\infty}du\, e^{-s u^2-2\pi i n u} \nonumber \\
&= \frac{\sqrt{\pi}} {\Gamma(\Delta)} \int_0^\infty ds\,s^{\Delta-\frac32} \exp\left[-x^2s-\frac{\pi^2 n^2}{s}\right]\ .
\end{align}
The zero mode should be treated separately. For $n=0$,
\begin{equation}
\widehat f(0) = \frac{\sqrt{\pi}}{\Gamma(\Delta)} \int_0^\infty ds\, s^{\Delta-\frac32}e^{-x^2s}
= \frac{\sqrt{\pi}\,\Gamma\!\left(\Delta-\frac12\right)} {\Gamma(\Delta)} x^{1-2\Delta}\ .
\end{equation}
For $n\neq 0$ we use a modified Bessel function of the second kind
\begin{equation}
\int_0^\infty ds\,s^{\nu-1}e^{-a s-b/s} = 2\left(\frac{b}{a}\right)^{\nu/2} K_\nu(2\sqrt{ab})\ .
\end{equation}
In the present case
\begin{equation}
\nu=\Delta-\frac12\,,\qquad a=x^2\,,\qquad b=\pi^2n^2\ .
\end{equation}
Therefore,
\begin{equation}
\widehat f(n)
=
\frac{2\sqrt{\pi}}{\Gamma(\Delta)}
\left(\frac{\pi |n|}{x}\right)^{\Delta-\frac12}
K_{\Delta-\frac12}(2\pi |n|x)\ .
\end{equation}
Putting the zero mode and the nonzero modes back into the Poisson formula gives
\begin{equation}
g^{\rm GFF}_{\Delta}(\tau,x) = \widehat f(0) + \sum_{n\neq 0} e^{2\pi i n\tau}\widehat f(n)\ .
\end{equation}
Since $\widehat f(n)=\widehat f(-n)$, the nonzero modes combine into cosines. Thus
\begin{equation}
g^{\rm GFF}_{\Delta}(\tau,x) =
\frac{\sqrt{\pi}\,\Gamma\!\left(\Delta-\frac12\right)} {\Gamma(\Delta)} x^{1-2\Delta}
+
\frac{4\sqrt{\pi}}{\Gamma(\Delta)} \sum_{n=1}^{\infty} \left(\frac{x}{\pi n}\right)^{-\Delta+\frac12} \cos(2\pi n \tau)\, K_{\frac12-\Delta}(2\pi n x)\ .
\end{equation}
Let us finally comment on the continuation in $\Delta$. The derivation above was performed for $\Delta>1/2$, where the original image sum is
absolutely convergent. The right-hand side, however, is meromorphic in
$\Delta$. The nonzero Fourier modes are entire functions of
$\Delta$, while the zero mode contains the factor $\Gamma\!\left(\Delta-\frac12\right)/\Gamma(\Delta)$. For $x>0$, the large-$n$ behavior of the Bessel function is exponentially
small,
\begin{equation}
K_{\frac12-\Delta}(2\pi n x)\sim e^{-2\pi n x}\,, \qquad n\rightarrow \infty\,,
\end{equation}
up to powers of $n$. Therefore, the nonzero-mode series converges locally
uniformly in $\Delta$. The Bessel representation consequently gives the
analytic continuation of the GFF image sum away from the poles of the zero
mode. In particular, it is valid in the region $|\operatorname{Im}\tau|<|x|$ used in the main text.

\section{Zero-frequency correlator}
\label{app: zero frequency}

In this appendix, we give more details on the construction of the zero Fourier mode $g^{(0)}(x)$. The full two-point function is obtained by solving the Klein-Gordon equation
\begin{equation}
    \left(\Box-\Delta_\phi(\Delta_\phi-4)\right)\Phi=0\,,
\end{equation}
on the Euclidean planar black hole background with the metric
\begin{equation}
ds^2=\frac{1}{z^2}
\left(\frac{dz^2}{f(z)}+f(z)d\tau^2+d\vec{x}^{\,2}\right)\,,   \qquad f(z)=1-z^4\ .
\end{equation}
Our conventions are the same as in the main text: the horizon sits at $z_H=1$, so that the inverse temperature is $\beta=\pi$. The boundary conditions for $\Phi$ may be looked up in \cite{Parisini:2023nbd,Buric:2025anb}.
\smallskip

Since the coordinate $\tau\sim\tau+\beta$ is periodic, the function $\Phi$ may be expanded in a Fourier series and we are after the zero Fourier mode $\Phi_0$. Furthermore, we are considering rotationally invariant solutions, so that $\Phi_0=\Phi_0(z,x)$. The Klein-Gordon operator commutes with $\partial_\tau$ and thus descends to a well-defined operator on the space of $\tau$-independent functions. Therefore, to complete the formulation of the problem, we need to write the boundary conditions satisfied by $\Phi_0$. To this end, we start with the Euclidean AdS bulk-to-boundary propagator
\begin{equation}
G_{\rm AdS}(z,\tau,x) = \frac{z^{\Delta_\phi}} {\left(\tau^2+x^2+z^2\right)^{\Delta_\phi}}\,,
\end{equation}
and make it thermal by summing over images along the Euclidean time circle,
\begin{equation}\label{thermal-AdS-propagator}
G_{\rm tAdS}(z,\tau,x)
=\sum_{n\in\mathbb{Z}}\frac{z^{\Delta_\phi}}{\left[(\tau+n\beta)^2+x^2+z^2\right]^{\Delta_\phi}}\ .
\end{equation}
The zero Fourier mode of the thermal propagator therefore reads
\begin{align}
&G_{\rm tAdS}^{(0)}(z,x) = \frac{1}{\beta}\int_0^\beta d\tau\,G_{\rm tAdS}(z,\tau,x)\\
& = \frac{z^{\Delta_\phi}}{\beta} \int_{-\infty}^\infty \frac{dt}{\left(t^2+x^2+z^2\right)^{\Delta_\phi}} = \frac{\sqrt{\pi}}{\beta}
\frac{\Gamma\left(\Delta_\phi-\frac12\right)}
{\Gamma(\Delta_\phi)}\frac{z^{\Delta_\phi}}{\left(z^2+x^2\right)^{\Delta_\phi-\frac12}} =
\frac{2}{\pi}
\frac{z^{3/2}}{z^2+x^2}\ . \nonumber
\end{align}
We have used that the image sum in \eqref{thermal-AdS-propagator} unfolds the integral over the circle into an integral over the real line. In the final step, we substituted the values $\Delta_\phi=3/2$ and $\beta=\pi$. In summary, we are solving the equation
\begin{equation}
    \left(\Box-\Delta_\phi(\Delta_\phi-4)\right)\Phi_0=0\,,
\end{equation}
subject to regularity in the bulk and the boundary condition
\begin{equation}\label{bc-s-Phi-appendix}
    \left. \Phi_0\right|_\partial = \left. G_{\rm tAdS}^{(0)}\right|_\partial\ .
\end{equation}
To write the boundary conditions more explicitly, we pass to the function $H(z,x)$, defined via the relation
\begin{equation}
\Phi_0(z,x) = G_{\rm tAdS}^{(0)}(z,x)+z^{3/2}H(z,x)\ .
\end{equation}
Furthermore, we introduce a coordinate $R$ by
\begin{equation}
x=\frac{R}{1-R^2}\,, \qquad R\in[0,1]\,,
\end{equation}
so that $R=1$ corresponds to spatial infinity. The equation is thus solved on the rectangular domain $(z,R)\in[0,1]^{2}$, subject to boundary conditions
\begin{equation}\label{H-bdy-conditions}
    H\big|_{R=1} = \partial_R H\big|_{R=0} = \partial_z H\big|_{z=0} = 0 \ .
\end{equation}
Regularity at the Euclidean horizon $z=1$ is imposed by evaluating the equation of motion at the horizon, where the coefficient of
$\partial_z^2 H$ vanishes. This gives the Robin-type condition
\begin{equation}\label{H-horizon-condition}
\left.\partial_z H\right|_{z=1} =
\frac14
\left[\left.\left(\partial_x^2+\frac{2}{x}\partial_x-\frac94\right)H\right|_{z=1}
-\frac{9x^4-22x^2+1}{2\pi(1+x^2)^3}\right]\ .
\end{equation}
The finite part of the zero-frequency correlator is then read off directly from the boundary value of $H$ according to
\begin{equation}
g^{(0)}(x)-\frac{2}{\pi x^2} = H(0,x)\ .
\end{equation}

\subsection{Alternative view on logarithmic terms}

We now explain a way, alternative to the one used in the main text, to determine the logarithmic terms in the short-distance expansion of the finite zero-frequency correlator. In particular, we derive the coefficient of the leading $x^2\log x$ term and describe the general procedure for obtaining the coefficient of $x^{2k}\log x$.

To begin with, we write the differential equation for $H(z,x)$
\begin{equation}\label{eq:zero-mode-H-pde}
\left[ (1-z^4)\partial_z^2 - 4z^3\partial_z - \frac94 z^2 +\partial_x^2 + \frac{2}{x}\partial_x\right]H
= \frac{z^2\left(9x^4-22x^2z^2+z^4\right)}{2\pi\left(z^2+x^2\right)^3}\ .
\end{equation}
Introduce polar variables on the local four-dimensional space,
\begin{equation}
\rho^2=z^2+x^2\,, \qquad \mu=\frac{z}{\rho}\ .
\end{equation}
In these coordinates, the right-hand side of \eqref{eq:zero-mode-H-pde} depends only on $\mu$
\begin{align}\label{source-S0}
S_0(\mu) \equiv \frac{z^2\left(9x^4-22x^2z^2+z^4\right)}{2\pi\left(z^2+x^2\right)^3}
&=\frac{1}{2\pi}\left(9\mu^2-40\mu^4+32\mu^6\right)\\
& = -\frac{1}{8\pi} U_0(\mu)-\frac{3}{8\pi} U_2(\mu) +\frac{1}{4\pi} U_6(\mu)\ . \nonumber
\end{align}
In the second line, we have rewritten the result in terms of the axisymmetric harmonics on $S^3$, i.e. Chebyshev polynomials $U_\ell(\mu)$. This form is relevant because the operator on the left-hand side of \eqref{eq:zero-mode-H-pde} takes the form $\mathcal{D}_- + \mathcal{D}_+$, with
\begin{equation}\label{form-of-operator}
\mathcal{D}_- = \partial_z^2 + \partial_x^2+\frac2x\partial_x\,, \qquad \mathcal{D}_+ = - z^2\left(z^2\partial_z^2 + 4z\partial_z + \frac94\right)\ .
\end{equation}
Upon passing to coordinates $(\rho,\mu)$, we see that $\mathcal{D}_-$ has $\rho$-degree $-2$, i.e. it maps functions of the form $\rho^n f(\mu)$ to those of the form $\rho^{n-2} g(\mu)$. Similarly, the operator $\mathcal{D}_+$ has $\rho$-degree $+2$. We look for solutions that have no negative $\rho$-degrees in the near-boundary expansion. To generate the source \eqref{source-S0}, which is of $\rho$-degree zero, one should act with $\mathcal{D}_-$ on a function of degree two. Generally, we have
\begin{equation}
\mathcal{D}_-\left[\rho^p U_\ell(\mu)\right] = \left[p(p+2)-\ell(\ell+2)\right] \rho^{p-2} U_\ell(\mu)\ .
\end{equation}
Therefore, we see that the $\ell=2$ component of the source cannot be generated, as the corresponding eigenvalue vanishes, 
\begin{equation}
   p(p+2) - \ell(\ell+2)= 0\,, \qquad \text{for} \qquad p=2\,, \ \ \ell = 2\ .
\end{equation}
Consequently, the required local solution contains a logarithm,
\begin{equation}
\mathcal{D}_- \left[\rho^2\log(\rho)\,U_2(\mu)\right] = 6 U_2(\mu)\ .
\end{equation}
Thus, near $\rho=0$, the solution $H$ has the asymptotic expansion
\begin{equation}
H(\rho,\mu)
= H_{\rm reg}(\rho,\mu)
-\frac{\rho^2}{64\pi} U_0(\mu)-\frac{\rho^2\log(\rho)}{16\pi}
U_2(\mu) - \frac{\rho^2}{160\pi} U_6(\mu)+\cdots\,,
\end{equation}
where $H_{\rm reg}(\rho,\mu)$ denotes the regular homogeneous contribution, whose coefficients are fixed by the global boundary conditions. At the conformal boundary $\{z=0\}$ one has $U_2(0) = U_6(0)=-1$. Hence
\begin{equation}
H(0,x) =
H_{\mathrm{reg}}(0,x) + \frac{x^2}{16\pi}\log(x) -\frac{3x^2}{320\pi} + \ldots\ .
\end{equation}
It follows that the short-distance expansion of the finite zero-frequency correlator begins as
\begin{equation}\label{eq:zero-mode-short-distance}
g^{(0)}(x)-\frac{2}{\pi x^2} =
\gamma_0 +\frac{x^2}{16\pi}\log(x) +\gamma_1 x^2 +O(x^4,x^6\log x)\ .
\end{equation}
Higher logarithmic terms are obtained similarly. Numerically fitting the boundary data to an expansion including terms $x^{2k}\log x$ gives a leading logarithmic coefficient that converges to $1/(16\pi)$, in agreement with the preceding local analysis.

\section{Details about the exact PDE solution}
\label{app:exact-zero-frequency}

\subsection{Large-momentum expansion}
\label{app:Near-boundary expansion}

In this appendix, we derive the large-spatial-momentum expansion of the momentum-space zero-frequency correlator \eqref{asymptotic-expansion-tildeg}. The horizon-regular radial solution satisfies
\begin{equation}\label{eq:zero-mode-radial-large-k}
\left[(1-z^4)R_k'(z)\right]'-\left(k^2+\frac94 z^2\right)R_k(z)=0\ .
\end{equation}
Introduce the radial logarithmic derivative
\begin{equation}
\mathcal{K}(z,k) \equiv \frac{R_k'(z)}{R_k(z)}\ .
\end{equation}
Written in terms of $\mathcal{K}$, the equation \eqref{eq:zero-mode-radial-large-k} becomes a first-order nonlinear Riccati differential equation
\begin{equation}\label{eq:zero-mode-riccati}
(1-z^4)\left(\partial_z\mathcal{K} +\mathcal{K}^2\right) -4z^3\mathcal{K} -k^2 -\frac94z^2 =0\ .
\end{equation}
We seek a formal large-momentum expansion of the form
\begin{equation}\label{eq:zero-mode-riccati-expansion}
\mathcal{K}(z,k) \sim
k\,s_{-1}(z) + s_0(z) + \frac{s_1(z)}{k} + \frac{s_2(z)}{k^2} +\cdots\ .
\end{equation}
At leading order in $k$, equation \eqref{eq:zero-mode-riccati} gives
\begin{equation}
(1-z^4)s_{-1}(z)^2=1\ .
\end{equation}
The branch compatible with the decaying pure-AdS solution is
\begin{equation}
s_{-1}(z) = -\frac{1}{\sqrt{1-z^4}}\,, \qquad s_{-1}(0)=-1\,,
\label{eq:s-minus-one}
\end{equation}
so that $\mathcal{K}^{-1}(0,k)\sim-1/k$ for large $k$ (cf. \eqref{AB AdS}). Substituting \eqref{eq:zero-mode-riccati-expansion} into
\eqref{eq:zero-mode-riccati} gives the recursion relation
\begin{equation}\label{eq:zero-mode-sn-recursion-solved}
s_n=
\frac{(1-z^4)s_{n-1}'-4z^3s_{n-1}+(1-z^4)\displaystyle\sum_{j=0}^{n-1}
s_j s_{n-1-j}-\frac94\,\delta_{n1}z^2}{2\sqrt{1-z^4}}\,,
\qquad n\geq 0\,,
\end{equation}
where the sum is absent for $n=0$. The first few coefficient functions are
\begin{align}
&s_0(z) = \frac{z^3}{1-z^4}\,,& 
&s_1(z) = \frac{z^2(z^4+3)}{8(1-z^4)^{3/2}}\,,
\\ \nonumber
&s_2(z) = \frac{z(3+6z^4-z^8)}{8(1-z^4)^2}\,,& 
&s_3(z) = \frac{24+369z^4-18z^8+25z^{12}}{128(1-z^4)^{5/2}}\ .
\end{align}
Their boundary values are
\begin{equation}
s_0(0)=s_1(0)=s_2(0)=0\,, \qquad s_3(0)=\frac{3}{16}\ .
\end{equation}
The zero-frequency correlator is, therefore,
\begin{equation}
\widetilde{g}^{(0)}(k) = -\frac{4\pi}{\mathcal K(0,k)} = \frac{4\pi}{k}
\left[1-\frac{3}{16k^4}+O(k^{-8})\right]^{-1} = \frac{4\pi}{k}+\frac{3\pi}{4k^5}+O(k^{-9})\ .
\end{equation}
Continuing in the same way, one obtains any desired number of terms in the asymptotic expansion. The horizon boundary condition affects the expansion only through terms that are exponentially small in $k$ and therefore does not modify the inverse-power asymptotic series.

\subsection{Radial equation integration}
\label{app:Radial equation integration}

The coefficients appearing in the short-distance expansion \eqref{eq:exact-short-distance-expansion} are obtained from momentum integrals involving the ratio $R_k(0)/R_k'(0)$. In this appendix, we describe the numerical determination of this ratio and the subsequent evaluation of the momentum integrals.

\paragraph{Numerical solution of the radial equation.}

For each fixed value of $k$, we solve the zero-frequency radial equation \eqref{eq:zero-mode-radial-large-k} from the horizon to the conformal boundary. The regularity conditions at the horizon were spelled out in \eqref{horizon-regularity-R-Rprime}, and we recall them here for convenience of the reader:
\begin{equation}\label{app:horizon-regularity}
R_k(1) = 1\,, \qquad R_k'(1) = -\frac14\left(k^2+\frac94\right)\ .
\end{equation}
The coefficient multiplying $R_k''$ in \eqref{eq:zero-mode-radial-large-k} vanishes at the horizon. It is therefore convenient to start the numerical integration at $z=1-\varepsilon$, with $\varepsilon>0$, using a regular near-horizon expansion
\begin{equation}\label{app:horizon-expansion}
R_k(z) = 1+r_1(k)(1-z)+r_2(k)(1-z)^2+O((1-z)^3)\ .
\end{equation}
Substitution into \eqref{eq:zero-mode-radial-large-k} gives
\begin{equation}
r_1(k) = \frac14\left(k^2+\frac94\right)\,, \qquad 
r_2(k) = \frac{1}{16}
\left[\left(k^2+\frac{57}{4}\right)r_1(k)-\frac92\right]\ .
\end{equation}
The initial conditions at $z=1-\varepsilon$ are consequently
\begin{align}\label{app:radial-initial-data}
R_k(1-\varepsilon)
& = 1+r_1(k)\varepsilon+r_2(k)\varepsilon^2 +O(\varepsilon^3)\,,\nonumber\\
R_k'(1-\varepsilon)
&=-r_1(k)-2r_2(k)\varepsilon+O(\varepsilon^2)\ .
\end{align}
After integrating \eqref{eq:zero-mode-radial-large-k} to $z=0$, we extract
the ratio $R_k(0)/R_k'(0)$.

\paragraph{Momentum integrals.}

The momentum integrals are evaluated by separating the integration range into a numerical region and an asymptotic region. For an integral of the form
\begin{equation}
\mathcal{I}
=
\int_0^\infty dk\,F(k)\,,
\end{equation}
we introduce a finite cutoff $\Lambda$ and write
\begin{equation}\label{app:integral-splitting}
\mathcal{I} = \int_0^\Lambda dk\,F(k) + \int_\Lambda^\infty dk\,F(k)\ .
\end{equation}
The first term is computed numerically. At every momentum required by the
quadrature routine, the radial equation is solved independently and the corresponding value of $R_k(0)/R_k'(0)$ is evaluated. The second term is obtained analytically from the large-$k$ expansion of the integrand. We check convergence by varying $\varepsilon$, the momentum cutoff $\Lambda$, and the numerical integration tolerances.

\subsection{Higher expansion coefficients}\label{app:Higher expansion coefficients}

In this appendix, we illustrate how the coefficients $\gamma_i$ are computed by explicitly working out the cases of $\gamma_1$ and $\gamma_2$. To compute $\gamma_1$, one might naively expand the sine kernel in the
position-space correlator and obtain
\begin{equation}
\gamma_1^{\mathrm{naive}} = \frac{1}{3\pi} \int_0^\infty dk\,k^3 \left(1+k\frac{A(k)}{B(k)}\right)\ .
\end{equation}
This expression cannot be used directly because the integral diverges logarithmically. Indeed, at large momentum,
\begin{equation}
k^3\left(1+k\frac{A(k)}{B(k)}\right) = -\frac{3}{16k}+O(k^{-5})\ .
\end{equation}
The leading asymptotic term must therefore be treated separately. Its finite-part sine transform is
\begin{equation}
-\frac{2}{\pi x}\operatorname{FP}\int_0^\infty dk\,
\sin(kx)\left(-\frac{3}{16k^4}\right)
=
x^2\left(
\frac{\log x}{16\pi}
+\frac{\gamma_{\mathrm E}}{16\pi}
-\frac{11}{96\pi}
\right).
\end{equation}
The logarithmic divergence of the momentum moment is removed by adding
$\frac{1}{16\pi}\log\Lambda$. It follows that
\begin{equation}
\gamma_1
=
\lim_{\Lambda\to\infty}
\left[
\frac{1}{3\pi}
\int_0^\Lambda dk\,k^3
\left(1+k\frac{A(k)}{B(k)}\right)
+\frac{1}{16\pi}\log\Lambda
\right]
+\frac{\gamma_{\mathrm E}}{16\pi}
-\frac{11}{96\pi}\ .
\end{equation}
Numerically, we find
\begin{equation}
\gamma_1\simeq -0.01617472265814\ .
\end{equation}
To compute $\gamma_2$, one may try to expand the sine kernel in
\eqref{eq:exact-position-space-correlator}, obtaining
\begin{equation}
\gamma_2^{\mathrm{naive}}
= -\frac{1}{60\pi}\int_0^\infty dk\,k^5\left(1+k\frac{A(k)}{B(k)}\right)\ .
\end{equation}
This integral is quadratically divergent. Indeed, the large-momentum
expansion gives
\begin{equation}
k^5\left(1+k\frac{A(k)}{B(k)}\right) =-\frac{3}{16}k-\frac{2637}{512k^3}+O(k^{-7})\ .
\end{equation}
The leading divergence is therefore
\begin{equation}
-\frac{3}{16}\int_0^\Lambda dk\,k = -\frac{3}{32}\Lambda^2\ .
\end{equation}
It follows that the finite coefficient $\gamma_2$ is given by
\begin{equation}
\gamma_2
=-\frac{1}{60\pi} \lim_{\Lambda\to\infty}
\left[\int_0^\Lambda dk\,k^5
\left(1+k\frac{A(k)}{B(k)}\right)+\frac{3}{32}\Lambda^2\right]\ .
\end{equation}
Numerically, we find
\begin{equation}
\gamma_2 \simeq 0.0003249565648\ .
\end{equation}
Higher coefficients $\gamma_i$ are obtained similarly.

\bibliographystyle{JHEP}
\bibliography{bibliography} 

\end{document}